\documentclass[apj]{emulateapj}

\newcommand{\hst}{\textsl{HST}}
\usepackage{graphicx}
\usepackage{rotating}

\newcommand{\lsim}{\mbox{$_<\atop^{\sim}$}}
\newcommand{\gsim}{\mbox{$_>\atop^{\sim}$}}

\newcommand{\mum}{$\,\mu$m}

\newcommand{\arcm}{$^{\prime}$}
\newcommand{\arcs}{$^{\prime\prime}$}

\newcommand{\spitzer}{\textsl{Spitzer}}

\newcommand{\galex}{\textsl{GALEX}}

\slugcomment{Accepted to the Astrophysical Journal}

\def\fesc {$f_{\mbox{esc}}$}

\def\lsun{$L_{\odot}$}

\def\fesc {$f_{\mbox{esc}}$}

\def\arcs{\ifmmode {''}\else $''$\fi}
\def\arcm{\ifmmode {'}\else $'$\fi}
\def\Msun{M$_{\odot}$}
\def\Myr{\Msun yr$^{-1}$}

\shorttitle{Escape Fraction at $z \sim 0.7$.}
\shortauthors{Bridge et al.}

\begin{document}


\title{A Spectroscopic Search for Leaking Lyman Continuum at $z \sim 0.7^{\dagger}$}


\author{Carrie R. Bridge \altaffilmark{1}, Harry I. Teplitz \altaffilmark{2}, Brian Siana \altaffilmark{1}, Claudia
  Scarlata \altaffilmark{3}, Christopher J. Conselice \altaffilmark{4}, Henry C. Ferguson\altaffilmark{5}, Thomas M. Brown\altaffilmark{5}, Mara Salvato\altaffilmark{1,8}, Gwen C. Rudie \altaffilmark{1}, Duilia F. de Mello\altaffilmark{6,7}, James Colbert\altaffilmark{3}, Jonathan P. Gardner\altaffilmark{7}, Mauro Giavalisco \altaffilmark{5}, Lee Armus \altaffilmark{3}}

\altaffiltext{1}{California Institute of Technology, 220-6, Pasadena, CA 91125}
\altaffiltext{2}{Infrared Processing and Analysis Center, MS 100-22, Caltech, Pasadena, CA 91125}
\altaffiltext{3}{$Spitzer$ Science Center, California Institute of Technology, 220-6, Pasadena, CA 91125}
\altaffiltext{4}{University of Nottingham, Nottingham, NG7 2RD, UK}
\altaffiltext{5}{Space Telescope Science Institute, 3700 San Martin Drive, Baltimore, MD 21218}
\altaffiltext{6}{Department of Physics, Catholic University of America, 620 Michigan Avenue, Washington DC 20064}
\altaffiltext{7}{Astrophysics Science Division, Observational Cosmology Laboratory, Code 665, Goddard Space Flight Center, Greenbelt MD 20771.}
\altaffiltext{8}{Max-Planck institute for Plasma Physics \& Excellence Cluster Boltzmans strasse 2, Garching 86748 Germany}

\altaffiltext{${\dagger}$}{Based on observations made with the NASA/ESA {\textsl{Hubble Space Telescope}}, obtained at the Space Telescope Science Institute, which is operated by the Association of Universities for Research in Astronomy, Inc., under NASA contract NAS 5-26555.  These observations are associated with programs 11236.}

\email{bridge@astro.caltech.edu}

\begin{abstract}

We present the results of rest-frame, UV slitless spectroscopic
observations of a sample of 32 $z\sim$0.7 Lyman break galaxy (LBG) analogs in the
COSMOS field.  The spectroscopic search was performed with the Solar Blind Channel (SBC) on {\textsl{Hubble Space Telescope}}.
We report the detection of leaking Lyman continuum (LyC) radiation from an AGN-starburst composite.  While we find no direct detections of LyC emission in the remainder of our sample,  we achieve 
individual lower limits ($3\sigma$) of the observed non-ionizing UV to LyC flux density
ratios, $f_{\nu}$(1500\AA\,)/$f_{\nu}$(830\AA\,) of 20 to
204 (median of 73.5) and 378.7 for the stack.  Assuming an intrinsic
Lyman break of 3.4 and an intergalactic medium (IGM) transmission of
LyC photons along the line of sight to the galaxy of $85\%$
we report an upper limit for the  {\it relative} escape fraction in
individual galaxies of $0.02-0.19$ and a stacked $3\sigma$ upper limit of
$0.01$.    We find no indication of a relative escape fraction near unity as seen in some
LBGs at $z\sim3$. Our UV spectra achieve the deepest limits to date at any redshift for the escape fraction in
individual sources.  The contrast between these
$z\sim0.7$ low escape fraction LBG analogs with $z\sim3$ LBGs suggests
that either the processes conducive to high $f_{\mathrm{esc}}$ are not
being selected for in the $z\lsim1$ samples or the average escape fraction is
decreasing from $z\sim3$ to $z\sim1$.  We discuss possible mechanisms which could affect the escape of LyC photons.
\end{abstract}



\section{Introduction}
\label{introduction}
Reionization brought an end to the cosmic ``dark
ages,'' during which the Universe contained a mostly neutral
intergalactic medium (IGM). This transition is thought to be triggered by either the radiation from QSOs or from galaxies
containing large populations of massive stars.  Recent determinations of the high
redshift QSO luminosity function suggest that QSO space densities are
too low to reionize the universe at $z>6$
\citep{2010AJ....139..906W,2009ApJ...693....8B,2008ApJ...675...49S,2008AJ....135.1057J}, with the exception of \citet{2010ApJ...710.1498G}.  The contribution of galaxies to the ionizing UV
background \citep[][]{1999ApJ...514..648M} depends upon how many of
the Lyman continuum (LyC) photons produced by massive young stars escape a galaxy, evading absorption by neutral hydrogen atoms or dust grains.   Direct measurement of the escape fraction, \fesc, is impossible at the epoch of
reionization, because intervening absorbers make the IGM opaque to LyC
photons.  Instead, \fesc\ must be measured at lower redshifts
($z\lsim3.5$) in objects that are analogous to the galaxies
responsible for reionization. 

Detection of escaping LyC photons has eluded most surveys.  Observations of Lyman break galaxies (LBGs) at $z\sim 3$ \citep{2001ApJ...546..665S,2006ApJ...651..688S,2009ApJ...692.1287I} suggest that in $\sim$10\% of starbursts the escape fraction
is quite large, nearing unity.  In contrast, there are currently no LyC detections locally ($z\lsim1$), despite tremendous effort \citep[][see Siana et al. 2010 for a review]{1995ApJ...454L..19L,1997MNRAS.289..629G,2001A&A...375..805D,
 2003ApJ...598..878M,2007ApJ...668...62S,2009ApJ...692.1476C}.   One explanation suggested by several authors is that the cosmic average escape fraction 
evolves with redshift \citep{2006MNRAS.371L...1I,2007ApJ...668...62S,2010arXiv1001.3412S}.   One notable difference between high and low redshift studies of the escape fraction is the wavelength range used to measure the LyC flux.   All previous surveys searching for intermediate redshift LyC leaking galaxies have utilized broad-band photometry which probe $\sim 700$\AA\, while $z\sim3$ studies measure the LyC just below the Lyman limit.   Probing these shorter wavelengths increases the sensitivity to the star formation history, dust extinction and IGM absorption.  This means a decrease in star formation within the last 10 Myr would significantly lower the flux at 700\AA\, compared to 900\AA\,, weakening the LyC limits measured from broad-band photometry.   It is with this difference in mind that we have undertaken a large spectroscopic program with the {\it Hubble Space Telescope} (\hst)
to study the escape fraction in luminous starbursts at $z\sim 0.7$  (GO 11236; PI:Teplitz).  

Using the Solar Blind Channel (SBC)  of the Advanced Camera for Surveys (ACS), we have obtained far-ultraviolet (FUV) spectra of 32
starbursts chosen to be good analogs of LBGs. The observations presented in this paper cover rest-frame $\sim$750-1100 \AA\,, and provide the first spectroscopic search for LyC photons from galaxies at $0<z<3$, as well as the deepest limits on the escape fraction for individual galaxies at any redshift.  
Spectroscopy has an advantage over
broad-band imaging by being able to probe closer to the Lyman limit
(up to 880\AA\ versus 700\AA\,), while being on-board \hst provides high spatial resolution making it possible to study any offset
of LyC radiation compared to UV emission.  The large sample size provides enough number 
statistics to make a meaningful comparison with the rare, high-z, LBG detections.  

While no LyC flux is conclusively detected (excluding the detection in an AGN-starburst galaxy), we use these spectra to 
place strict limits on the UV ($\sim$1025\AA) to 
LyC ($780-860$\AA) ratio and infer the limit on the escape fraction of ionizing photons.  
  In Section 2, we describe the selection of LBG-analog targets; in Section 3,
we present the observations and data reduction.  The method of measuring
limits on the escape fraction is presented in Section 4.  In Sections 5 and 6, we give the results
and discuss their implications.  A cosmology of $\Omega_{\rm M}=0.3, \Omega_\Lambda=0.70$, and
$H_{0}$=70\,km\,s$^{-1}$\,Mpc$^{-1}$ is assumed throughout.

\section{Sample Selection}
\label{sec:selection}

Our aim was to chose galaxies at moderate redshift that are analogs to the $z\sim 3$\ LBG population, which is conventionally taken
as accessible analogs to $z>6$\ galaxies. Probing $z\sim1$ analogs provides the opportunity for a wide range of observations,
together with the opportunity to study the possible evolution of the escape fraction 
with redshift.  \citet{2007ApJS..173..441H} 
demonstrated the similarity between LBGs and the compact subset of local 
UV luminous galaxies \citep[UVLG;][]{2005ApJ...619L..35H}.  These sources
share many properties with LBGs including luminosity, size, specific
star formation rate (SFR), mass, and metallicity \cite{2005ApJ...619L..35H}.  The compact UVLG population is
quite rare locally (about 1 per square degree at $z<0.3$), but its numbers
increase with redshift.  They are identified by their UV luminosity
($L>2\times10^{10}$\ \lsun) and surface brightness ($I_{FUV} > 10^8$\ \lsun\
kpc$^{-2}$) \cite{2007ApJS..173..441H} .

The wavelength range of the SBC/PR130L configuration limits the redshift
range in which we can measure the LyC to $0.6<z<0.85$ .  
Furthermore, SBC prism spectra are
slitless, introducing a smearing effect, whereby the spectral resolution is degraded 
proportional to the spatial extent of the target.  Thus, we further restricted
our selection to compact galaxies, which is in line with the LBG analog selection.

To choose appropriate targets, we needed access to deep UV photometry,
good redshift estimates, and a measure of compactness and surface
brightness.  These three datasets are available in the COSMOS field \citep{2007ApJS..172...38S}.    We derived the UV luminosity from {\textsl{Galex Evolution Explorer}} (\galex) photometry \citep{2007ApJS..172..468Z}, the size from \hst (F814W)
imaging \citep{2007ApJS..172..196K} and an accurate photometric redshift with the COSMOS 30+ band photometry \citep{2007ApJS..172..117M}.

\begin{figure}
  \centering
  \includegraphics[width=100mm]{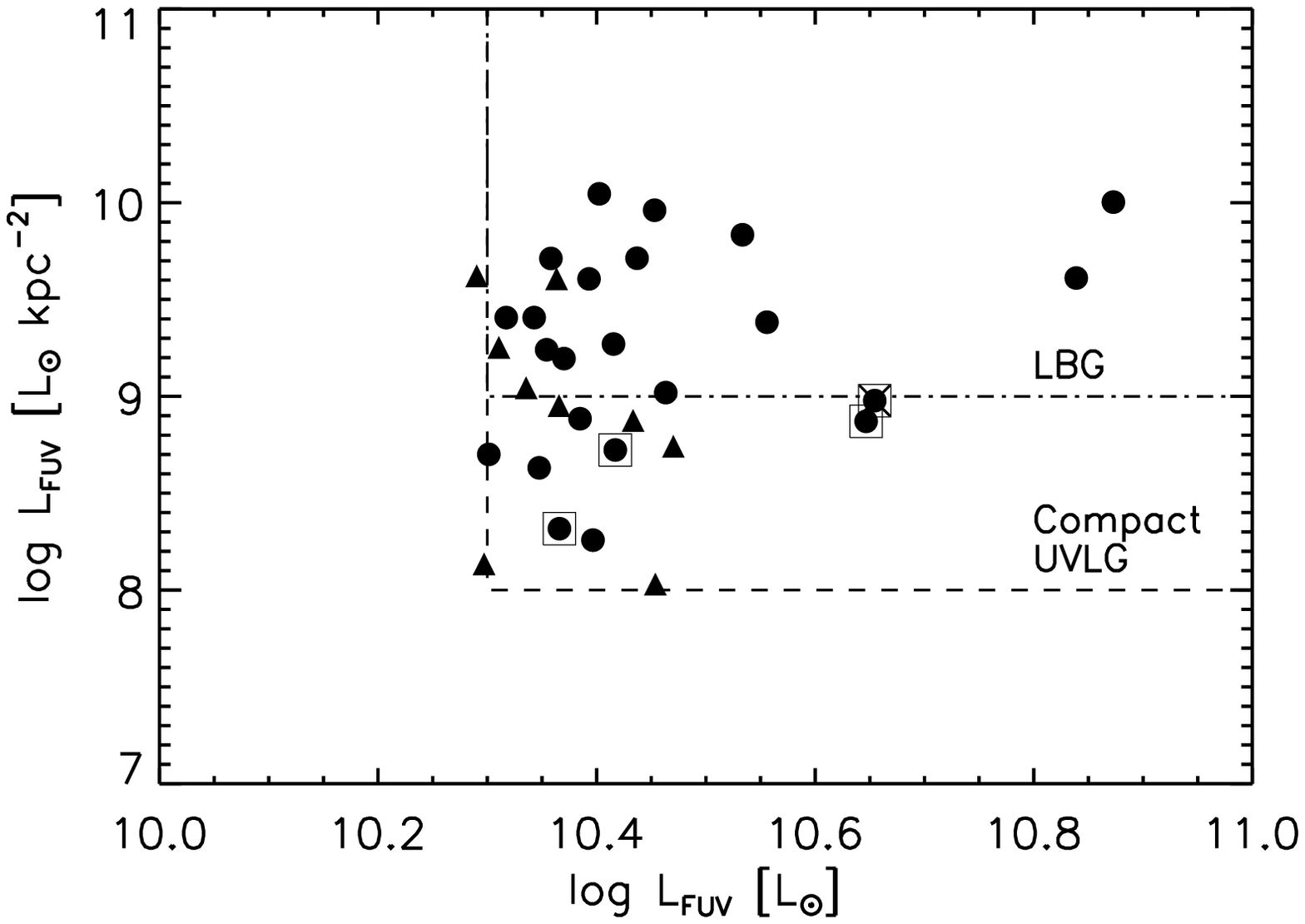}
\caption[surf_brightness]{ FUV luminosity vs.  FUV surface
  brightness for our sample of 32 galaxies (circles and triangles) with
  $0.65<z<0.85$\ in COSMOS.  The dash-dotted line indicates
  the region occupied by $z\sim 3$\ LBGs \cite{2007ApJS..173..441H,2002ARA&A..40..579G}, while the dashed line
  marks the region occupied by local compact UV luminous galaxies,
  thought to be analogs for $z\sim3$ LBGs \cite{2007ApJS..173..441H}.  The
  triangles highlight the galaxies that were not detected in the SBC prism observations. The open
  squares indicate galaxies with no ACS imaging, and the ``x'' (within a square) notes the
  target found to be an AGN--starburst composite (C-UVLG-17). Note that the angular sizes of our galaxies are derived using rest-frame optical ACS images, which are typically larger than the sizes measured by rest-frame UV images as in the case of \citet{2007ApJS..173..441H}. }
\label{fig:surf_bright}
\end{figure}

We selected non-point sources with large $L_{FUV}$\ and $I_{FUV}$\
analogous to LBGs (Figure \ref{fig:surf_bright} ).  After visual inspection we discarded 15 sources for technical reasons not
obvious in the catalog, such as a close proximity to the edge of the \galex\ field of view which suffers from artifacts.  There are 32 sources remaining which meet our
criteria.  These sources are the brightest UV galaxies at $z\sim 0.7$,
but their luminosity (log $L_{UV} \gsim 10.4$\ \lsun) is similar to that of
typical LBGs at $z\sim 3$.  An $\sim L*$\ galaxy at $z>6$\ would fall at the faint
end of the same range.  The UV luminosity of the targets corresponds
to a SFR of 8--45 \Myr, without correcting for dust extinction.

\begin{figure}
  \centering
  \includegraphics[width=130mm]{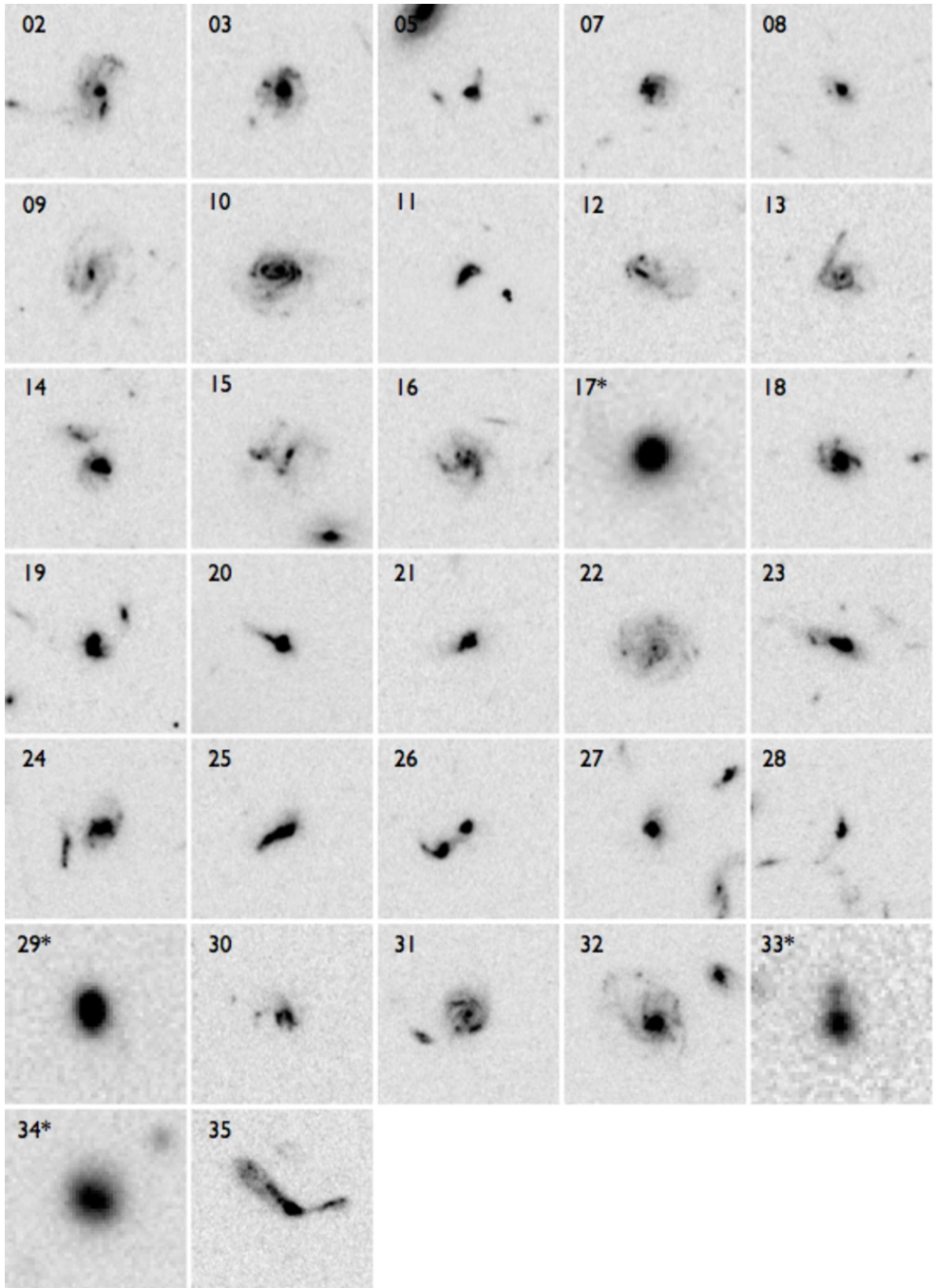}
\caption[montage]{Optical images of the 32 $z\sim0.7$ LBG
  analogs. Thumbnails are F814W \hst\ images except those marked with an asterisk  (nos. 17,29,33,34) which are CFHT ground-based $i'$.  The images are $7''$ on a side, with the object ID noted in the left corner. }
\label{fig:montage}
\end{figure}

The 32 selected galaxies (Figure \ref{fig:montage}) share many observable properties with LBGs.
By inference, they are the rare, young, strongly star-forming objects
at moderate redshift that have just reached an evolutionary stage
which was common at $z\sim 3$.  

\section{Observations and Reductions}
\label{obser and reduc}

\subsection{Slitless UV Spectroscopy}
\label{uv spec}

The SBC uses a  Multi-Anode Microchannel Array (MAMA)
that has no read noise and is not sensitive to cosmic rays.  Targets were positioned to land on the lower
right quadrant of the detector to minimize the dark current ``glow'', which was noted by \citet{2006AJ....132..853T} to increase with exposure time.

The SBC field of view is $31''\times 35''$, with a  plate scale of 0\farcs32 pixel$^{-1}$.  The SBC PR130L, which covers 1250 - 2000\AA\ with
a variable dispersion from 1.65\AA\ pixel$^{-1}$ at 1250\AA\ to
20.2\AA\ pixel$^{-1}$ at 1800\AA, is used because it is
not sensitive to the Ly$\alpha$ airglow which dominates the
background in the PR110L prism.  The PR130L prism is still
sensitive to other airglow lines so the observations were taken with the
SHADOW constraint reducing the airglow by a factor of 10.  

\begin{figure}
  \centering
  \includegraphics[width=130mm]{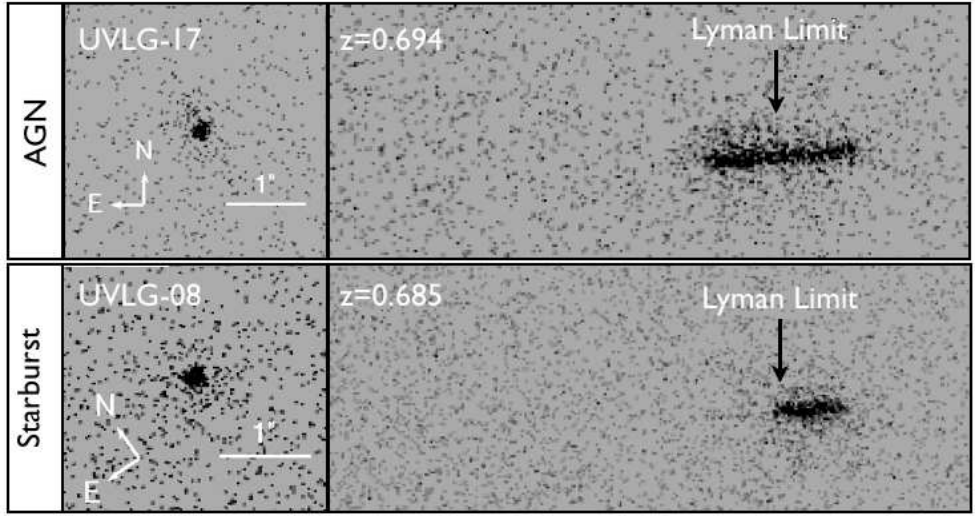}
\caption[2D Spectrum of C-UVLG-17 and UVLG-08]{Example of the SBC F150LP direct UV
  image (left) and two-dimensional spectra (right)
of an AGN (UVLG-17) and a starburst galaxy (UVLG-08) in our
sample.  The Lyman limit is noted in each spectra.  Flux below 912\AA\
is clearly detected in the AGN, while no LyC detection is found in the
starburst galaxy.  Wavelength increases from the left to right
covering $\sim$1220-2000\AA\ observed.}
\label{fig:2dspect}
\end{figure}

\begin{figure}
  \centering
  \includegraphics[width=150mm]{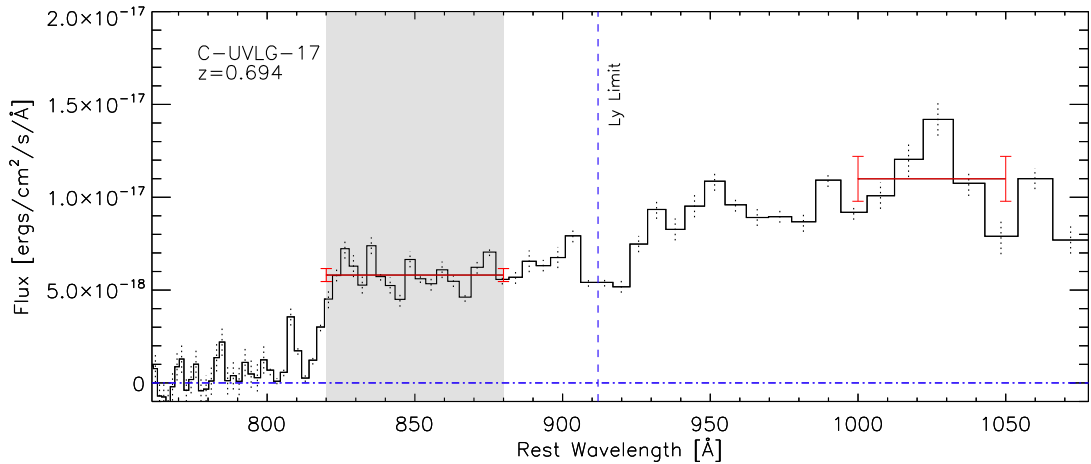}
\caption[Spectrum of C-UVLG-17]{Individual spectrum of galaxy
  C-UVLG-17.  Below the Lyman limit UV flux is clearly detected,
  showing that if LyC photons were escaping at the $\sim15\%$ level it would be detected by our observations.  The source of these photons is likely the Type 2 AGN.}
\label{fig:agn17}
\end{figure}

The sources have a median \galex\ near-UV magnitude of 22.5
(AB), and required 1-4 \hst\ orbits in the shadow position (1500s/orbit; see Table \ref{tab:result} for details).  Direct imaging (two
480s exposures/orbit) of our targets at 1600\AA\, was
also acquired in the same visit with the F150LP filter to establish a zero point of the
wavelength scale.  These data were taken over
2007 December - 2009 January in Cycle 16.  An example of the direct SBC image and corresponding two-dimensional spectra is shown in Figure \ref{fig:2dspect}
 for a typical starburst galaxy in our sample and the target which is an AGN-starburst composite (also see Figure \ref{fig:agn17} and Section \ref{subsec:agn} for a discussion).

The flat-fielding and dark subtraction of the SBC data were performed  by the \hst\, pipeline.  Since this work relies on the measurement of flux below
the Lyman limit, and our observations are background limited sky subtraction must be done with care.  The mean value for a region of sky, typically
300 pixels $\times$ 150 pixels above and below the target, is measured for
each prism exposure.  The average sky 
background in each individual exposure is subtracted from each flat-fielded prism
image.  The spectra were extracted in PyRAF using the {\it aXe}
slitless spectroscopy reduction package
\citep[ver. 1.6][]{2006hstc.conf...79W,2006hstc.conf...85K}, specifically
designed for \hst\ grism and prism data.  

In the extraction, the position of the galaxy on the direct image is used in
combination with the header dither parameters to locate the
spectrum on the dispersed image and assign a wavelength to
each pixel in the spectrum.  The spectral trace and wavelength
solutions are defined with respect to a reference position, which is
measured by running SExtractor \citep{1996A&AS..117..393B} on a direct
image for each orbit.  A master catalog was generated for each set of
prism images taken in a single orbit.  Using the master catalogs, the
{\it aXe} software generates pixel extraction tables (PETs), which contain a
spectral mapping for each pixel in the spectrum.  Various sizes for the extraction window were explored, with an extraction width approximately the size of the object maximizing the signal to noise of the spectra. The signal to noise was further increased (a factor of $\sim$1.4) by using {\it aXe's} optimal weighting extraction, which weights the pixels as a function of distance from the spectral trace rather than giving each pixel the same weight when generating the one dimensional spectra.
The wavelength solution over 1300-1700\AA\, is accurate to within a few angstroms.  The spectra were flux calibrated using the standard sensitivity curve for the PR130L prism \citep{2006acs..rept....3L}.  

The flux and wavelength-calibrated spectra for each prism image were
co-added weighting each exposure by the rms of the background in each individual exposure.  This was done to account for increasing dark current with subsequent exposures.  For an individual target the dark current was a factor of 4 higher in the forth orbit compared to the first.   The uncertainty in the spectral flux was estimated differently for the wavelength regions redward and blueward of the Lyman limit.  Blueward of 912\AA, where we typically measured no signal (i.e., no detection of LyC)  we are essentially probing the background.  We extracted regions of blank sky and found that 
the standard deviation of the sky flux between the observed wavelengths typically used to measure the LyC in our $z\sim0.7$ sample was comparable to the standard deviation of the flux in our galaxies. We therefore use the standard deviation of the flux between 780 and 860\AA\ rest frame (the region used to measure LyC flux limits) as a conservative estimate of the uncertainty. However, redward of the Lyman break, where flux is detected, the errors at each wavelength bin in the final one-dimensional co-added spectra are the $1\sigma$ $weighted$ standard deviation since there were larger.   The final one-dimensional spectra are presented in Figure 5.

\subsection{Optical Spectroscopy}
\label{sec:opt_spec}

Accurate spectroscopic redshifts are required to identify
the observed wavelengths corresponding to the
rest-frame Lyman limit and LyC.   Redshifts
cannot be measured from the FUV spectra themselves since the prism probes blueward of
Ly$\alpha$ in the rest frame and the resolution is too poor for the identification of absorption lines.  The photometric redshifts, although sufficient for selection purposes, have an uncertainty corresponding to $\sim$200\AA\, at $z\sim0.7$. Optical spectroscopy for six of our targets were taken from the public zCOSMOS program\footnote{Based on zCOSMOS observations carried out using the Very Large Telescope at the ESO Paranal Observatory under program ID: LP175.A-0839} \citep{2009ApJS..184..218L}.  An additional 23 spectra were obtained at the Palomar Hale 5 m Telescope using the Double Spectrograph \citep[hereafter Double Spec;][]{1982PASP...94..586O} in 2008 and 2009 February with seeing conditions ranging from 1\arcs\, to 2\farcs5 for the
majority of the observations and light cirrus.  
This instrument uses a pair of CCD cameras which simultaneously obtain long-slit spectra over a ``blue'' range of 3500- 5600\AA\, and a ``red'' range of 6000-8500\AA\,.  The
observing strategy consisted of acquiring a series of 600-900s exposures with a slit width of 1\farcs5.  We obtained a total integration time of 40-90 minutes for each of the 23 galaxies.

The data were reduced using standard IRAF tasks and wavelength calibrated with He, Ne, and Ar reference lamps.  The spectral features used for the redshift
determination were typically a combination of O[II] at 3727\AA\,, O[III] at 5007\AA\, and in some
cases H$\beta$ at 4861\AA\,.    The spectroscopic redshifts confirmed that all the observed galaxies were within the $0.6<z<0.85$ range needed to
measure the escape fraction with the UV observations.  The mean spectroscopic redshift of the sample is $z=0.679$.  Due to poor observing conditions, three targets remain with no spectroscopic redshift confirmation.  Since only one of these (C-UVLG-15)  had a detectable UV spectrum and the rest of the sample had photometric redshifts within a few percent of the spectroscopic redshift we assumed the photometric redshift for this galaxy.

\section{Deriving The Escape Fraction}
\label{esc_frac}

Typically, the escape fraction of LyC photons is measured relative to the number of photons escaping at $\lambda_{\mathrm{rest}}=$1500\AA\ \citep{2001ApJ...546..665S}.   This allows a straightforward calculation of the escape fraction with only two flux measurements (at 1500\AA\, and in the LyC) using the following equation:

\begin{equation}
\label{eqn: fesc_rel}
f_{\mathrm{esc,rel}} = \frac{(f_{1500}/f_{\mathrm {LyC}})_{\mathrm{int}}}{(f_{1500}/f_{\mathrm{LyC}})_{\mathrm{obs}}\times(S)}\times exp(\tau_{\mathrm{IGM,LyC}}),
\end{equation} 
 
\noindent where $(f_{1500}/f_{\mathrm{LyC}})_{\mathrm{int}}$ and $(f_{1500}/f_{\mathrm{LyC}})_{\mathrm{obs}}$ are the
intrinsic and observed LyC flux
density ratios.  The ``LyC'' refers to the wavelength at which LyC is
 measured (780-860\AA\ for this study), $S$ is the scaling
factor discussed below and $\tau_{\mathrm{IGM,LyC}}$ is the IGM optical depth
for LyC photons along the line of sight to the galaxy. 
The intrinsic drop between the rest-frame FUV
(1500\AA\,) and the LyC (700-900\AA\,) is highly uncertain
and cannot be observed.  The assumed value of $f_{\nu}$(1500\AA\,)/$f_{\nu}$(LyC)
has varied by study from 3
\citep{2001ApJ...546..665S,2006ApJ...651..688S} to 6 or 8
\citep{2007ApJ...668...62S}.    

Since we probe LyC radiation at slightly bluer wavelengths, at 830\AA\ compared with \citep{2001ApJ...546..665S,2006ApJ...651..688S}, we used the spectral energy distributions (SEDs) of \citet{2003MNRAS.344.1000B} and estimated
the break amplitude for $f_{\nu}$(1500\AA\,)/$f_{\nu}$(830\AA\,) to be 3.4 based on $f_{\nu}$(1500\AA\,)/$f_{\nu}$(900\AA\,)$=3$. 
We assume a factor of $\sim$1.2 ($exp(\tau_{\mathrm{IGM,LyC}})$) reduction in $f_{\mathrm{LyC}}$/$f_{1500}$ for the neutral hydrogen opacity in the IGM, modeled in the same manner as \citet{1995ApJ...441...18M} and \citet{2007ApJ...668...62S}.   The SBC spectra measure up to rest frame $\sim1080$ \AA\ for our targets.
Therefore, to calculate the escaping UV photons at 1500\AA\,, we use the flux within our
spectrum between rest frame 1000 and 1050\AA\ and apply a scaling factor ($S$) to estimate
the flux at 1500\AA\ ($f_{1500}$).  The sources in our sample are by selection, blue objects
with low to moderate levels of extinction.  
Using the SEDs of \citet{2003MNRAS.344.1000B}, and assuming 0.4 solar metallicity, constant star formation with a
300 Myr old population, we derive the scaling factor to go from the $f_{1025}$
to $f_{1500}$ to be 1.5 in $f_{\lambda}$.   The 1000-1050\AA\ flux measurement does not consider the possible effect, if any, of the Ly$\beta$ $\lambda$1026 absorption line as seen in a composite spectrum of $z\sim3$ LBGs \cite{2006ApJ...651..688S}.   Although the spectral resolution is too low to estimate the strength of this line, it likely has little impact since our measure of the continuum flux is averaged over 50 \AA\.  The fluxes derived from the spectra are also consistent with aperture photometry derived using the F150LP images. 

As we mentioned earlier, due to the slitless nature of the spectra there is spectral smearing along the dispersion direction proportional to the spatial size of the object.  A different red wavelength cutoff was assumed for each object based on the galaxy's size in the SBC direct image to ensure that light redward of the Lyman limit did not contaminate the LyC flux measurement.  The sources have typical radii of  $\sim$7-20 pixels in F150LP corresponding to a red cutoff of range of $\sim$820-880\AA\,.  Furthermore, the sensitivity drops sharply at the blue end and therefore we consider only the regions at rest wavelengths $>780$\AA\,.  The final spectral region used when estimating the LyC flux is between $\sim$780 and 880\AA\, with the red cutoff changing as a function of galaxy size (shaded regions in Figure 5). 

The continuum is relatively flat around 850\AA\ for galaxies that are
actively forming stars \citep{2003MNRAS.344.1000B}, like those in our sample.  We therefore
integrate the observed spectrum over several resolution elements
to increase the signal-to-noise ratio.  The wavelength solutions of the SBC
are accurate to a few angstroms between observed 1300\AA\ and 1700\AA\
over the SBC field of view, and
the flux calibrations over this wavelength range are accurate to approximately
$5\%$ \citep{2006acs..rept....9L}.  The flux below the Lyman limit ($f_{830}$) is taken to be the average flux between 780 and 880\AA\, (again the red cutoff depends on the galaxy size). The uncertainty in
$f_{830}$ is derived using the following equation: 

\begin{equation}
\label{eq:1}
f_{er,830}=\frac{\sqrt{\sum(\Delta f_{er}^2\times\Delta\lambda^2)}}{\Delta\lambda_{\mathrm{tot}}},
\end{equation}

where $\Delta f_{er}$ is the standard deviation of the flux in each pixel, $\Delta\lambda$ is the size of each pixel in
angstroms (this value changes as a function of wavelength), and
$\Delta\lambda_{\mathrm{tot}}$ is the total wavelength range being averaged. 
The amount of escaping radiation
below the Lyman limit is typically reported using a {\it relative}
escape fraction (defined earlier in this section) or through a UV-to-LyC flux density ratio ($f_{\nu}$(1500\AA)$/f_{\nu}$(830\AA).  The former measure, however, requires an
assumption for the intrinsic Lyman break which is not well
constrained.  In the next section, we present both the UV-to-LyC flux density ratios and
the inferred relative escape fractions for completeness.

\section{Results}
\label{sec:results}

We find one detection of escaping LyC radiation in an AGN starburst composite, but no direct detections of far-UV flux 
in our remaining sample of 31 $z\sim0.7$ LBG analogs.   
 Nine galaxies (C-UVLG-3, 9, 10, 13, 14, 16, 22, 31, 32) were not detected in the
direct F150LP image or PR130L spectra.  Measuring limits on the LyC escape for these objects requires careful cross-calibration between \hst\ and \galex\ data due to the significant difference in angular resolutions (\galex\ point spread function (PSF) $\sim5$\arcsec in the FUV band) and will be presented in a future paper (H.I. Teplitz et al., 2010 in preparatio).  The non-detections are likely due to in part to their large size which resulted in UV surface brightnesses below the sensitivity of our observations.  Another likely contributing factor is that 7/9 of the undetected galaxies had the largest extinctions within our sample $E(B-V)>0.3$, based on SED fitting.

The observed flux density ratio, $f_{\nu}{1500}/f_{\nu}{830}$, in the individual sources range
from 20 to 204 with a median of 73.5 ($3\sigma$ lower limits). In order to
convert these ratios into a {\it relative} escape fraction, we apply
Equation (\ref{eqn: fesc_rel}) assuming an average
IGM transmission of 0.85 and a value of 3.4 or 7 for the intrinsic
Lyman break (see Table \ref{tab:result}).  Our far-UV
sensitivities give $f_{\mathrm{esc,rel}}$ close to zero, with individual $3\sigma$ upper limits
ranging from $0.01-0.19$.   Since we have assumed an average IGM transmission, these upper limits are likely to be even lower in the majority of these objects since the IGM opacity at low redshift is dominated by very few opaque lines of sight \cite{2007ApJ...668...62S}.

In order to increase our sensitivity further, we stacked the non-detections with  UV sizes $\lsim 0\farcs78$ in diameter, which corresponds to a red cutoff of $\sim$860\AA\,.  This red cutoff was chosen to maximizing the number of galaxies in the stack while probing as closely to the Lyman limit as possible.  The stack was composed of 18 galaxies placing a $3 \sigma$ lower limit on  $f_{\nu}{1500}/f_{\nu}{830}=$378.7 and a $3\sigma$ upper limit of $f_{\mathrm{esc,rel}}<0.01$ (Figure \ref{fig:stackedfull}) .  In addition to the global stack, we separated the sample by morphology, stacking the eight galaxies which were visually classified in the \hst\ F814W image to be undergoing a merger event and a radius of $\le0\farcs7$ (red cutoff of 865\AA\,).  We find $3\sigma$ lower limits for  $f_{\nu}{1500}/f_{\nu}{830}=$223.2 and a $f_{\mathrm{esc,rel}}<0.02$ ($3\sigma$ upper limit) for galaxy mergers (see Figure \ref{fig:stackedmerger} and Section \ref{sec:galaxy-mergers} for further discussion).

\subsection{LyC Emission: AGN and Star Formation?}
\label{subsec:agn}

Flux below the Lyman limit was detected in C-UVLG-17 (hereafter T17), however it is unclear what process is responsible for the FUV emission (Figure ~\ref{fig:agn17}).  T17 is detected in the X-ray with a luminosity of $\sim2\times10^{43}$erg s$^{-1}$ (2-10KeV) \citep{2010ApJ...716..348B}, and is a Type 2 AGN based on the optical spectrum \citep{2006ApJ...644..100P}.  Although it is highly probable that the LyC flux is coming from the AGN,  it is possible that some flux has its origin in young massive stars.  The COSMOS photometry, with 30-bands, was fit using a library of ``hybrid'' templates tuned for sources hosting an AGN even if hidden \citep[see][for details]{2009ApJ...690.1250S}.  The best-fit SED is a pure starburst, although the fit is poor.  The IRAC colors do not place T17 in the locus of optically bright unobscured AGN and is non-variable (over a five year period).  The spectrum shows no break at the Lyman limit, with ionizing flux detected down to $\lambda \sim 810$ \AA.  Below this wavelength, the continuum is strongly suppressed, indicative of a foreground, high-column density absorber at $z\sim 0.55$. 

Although this source harbors an AGN, it is worth noting that a large fraction of the LyC radiation could be from the starburst, with the AGN clearing lines of sight.  It is impossible to disentangle the fractional contribution of the LyC radiation from the two sources.   In the context of this paper we do not consider the LyC emission from this galaxy to be analogous to the LyC detections at high-$z$, although this type of LyC emission is interesting and possibly more common at higher redshift.

\subsection{Foreground Contamination}
\label{sec:contamination}

Higher redshift ($z\sim3$) studies have detected LyC photons in approximately $10\%$ of galaxies \citep{2006ApJ...651..688S,2009ApJ...692.1287I}.   
\citet{2010MNRAS.404.1672V} estimate that at $z\sim3$ there is a $50\%$ probability that 1/3 of the galaxies with detected LyC emission are the result of foreground contamination by low-redshift blue galaxies within $\sim$1\arcs of the high-redshift galaxy.  Figure \ref{fig:panels} highlights an example of foreground contamination found in C-UVLG-14 of our sample.  The NUV \galex\ magnitude which is used in our sample selection has a PSF of $\sim4$\arcs which encompasses both galaxies.

The high-resolution  \hst\, F814W image \citep{2007ApJS..172...38S} reveals two galaxies separated by $\sim$1\farcs2 ($<$10kpc), exhibiting irregular morphologies consistent with a close galaxy pair.  The southern galaxy (hereafter GalA) has two compact nuclei and evidence of long tidal tails.  The northern galaxy (hereafter GalB) exhibits multiple star formation knots, an asymmetric morphology and tidal tails.  The two primary UV bright knots that make up GalB are referred to as GalB1 and GalB2 as noted in Figure \ref{fig:panels}.  

Although GalA was the intended target in our program ($z_{\mathrm{spec}}$=0.772), the high resolution SBC direct F150LP image revealed that GalB dominated the NUV and  FUV flux.   The \spitzer\, 24\mum\, detection is centered on GalA corresponding to an IR luminosity of $L_{[8-1000\micron\,]}=2.3\times10^{11}L_{\odot}$ and a SFR of $\sim40M_{\odot}$ $yr^{-1}$ \citep{1998ARA&A..36..189K}. 
GalA was not detected in the FUV (likely due to its high dust content), however, the UV spectrum of GalB2 (Figure \ref{fig:interloper}) would imply a {\it relative} escape fraction, $f_{\mathrm{esc,rel}}$ of $30\%\pm6\%$ and a  $f_{\nu}$(1500\AA)$/f_{\nu}$(830\AA) $=$14.3, if it was at the same redshift as GalA.  Follow-up spectroscopy with Keck DEIMOS revealed that GalB was a low-redshift interloper at $z=0.19$.  The close ($\sim1$\arcs) projected separation of these galaxies is a good example of foreground contamination and could explain some the $z\sim3$ detections of \citet{2009ApJ...692.1287I} particularly the ones with a $\sim1$\arcs\ offset between the LyC emission and assumed LBG counterpart.

\section{Discussion}
\label{discussion}

The deep UV spectra presented here achieve the lowest individual limits
of the ionizing escape fraction at any redshift.  We confirm the results of
shallower studies that suggest low escape fractions in moderate
redshift starbursts.  
We now present our findings in the context of recent studies and discuss some possible mechanisms that could be responsible for
the apparent lack of $z\sim1$ starbursts with large escape fractions.

\subsection{Comparison with Recent Work}

Recently, \citet{2010arXiv1001.3412S} used \hst ACS/SBC 1500 \AA\ imaging of 15 $z\sim1.3$ starburst galaxies in the GOODS field, obtaining a stacked $3\sigma$ limit of $f_{\nu}(700)/f_{\nu}(1500) < 0.02$.  With the inclusion of previous studies \citep{2003ApJ...598..878M,2007ApJ...668...62S} they state that no more than 8\% of star-forming galaxies at $z\sim$1 have relative escape fractions greater than 0.50.   \citet{2009ApJ...692.1476C} stacked the \galex far-UV (1500 \AA) fluxes of a much larger sample (626 galaxies) and obtain a similar $3\sigma$ upper limit of $f_{\nu}(700)/f_{\nu}(1500)< 0.012$, confirming that starbursts at $z\sim1$ have low ionizing emissivities.  All of these studies have used broadband UV imaging to probe the LyC at $\sim700$\AA\, whereas the detections at $z\sim3$ have been obtained through spectroscopy \citep{2001ApJ...546..665S,2006ApJ...651..688S} and narrowband imaging \citep{2009ApJ...692.1287I} where the LyC is sampled just below the Lyman limit ($880-910$ \AA).

This is the first $z\sim1$ spectroscopic study to provide LyC measurements much
closer to the Lyman limit and is therefore more directly comparable to
these higher redshift studies.  \citet{2006ApJ...651..688S}, with a
sample of 14 LBGs probe a slightly
higher UV luminosity (see Figure 10) and detect two objects with large escape
fractions (with LyC-to-UV ratios significantly above our limits).  One
of these objects likely has a lower escape fraction than
initially reported as it was not detected in deep narrowband imaging
probing the LyC \citep{2009ApJ...692.1287I}.  Therefore,
only one of 14 objects has a significant detection.  In a stack of the 12 undetected sources, \cite{2006ApJ...651..688S} find an observed lower limit on the UV-to-LyC flux density ratio at $z\sim$3 of  $f_{1500}/f_{900}>43.0$, compared to our $z\sim1$ lower limit of $f_{1500}/f_{830}>379$.  Similarly,
\citet{2009ApJ...692.1287I}, which probe comparable rest-frame UV
luminosities as our $z<1$ sample, detect large escape fractions in
$\sim 10\%$ of LBGs and Ly$\alpha$ emitters at $z=3.1$.    If the LBG
analogs in our sample have ionizing properties similar to LBGs at
$z\sim3$, we would expect to detect approximately three objects with large
fractions of escaping LyC radiation, however, we detect none (excluding the AGN-starburst composite).   

Our individual limits are significantly better than the
studies at $z=3$, so if there are significant numbers of
galaxies with lower escape fractions (eg. $f_{\mathrm{esc,rel}}\sim 0.20$, rather than
unity), we would be able to detect them.  We see no evidence for this
scenario in our sample.  
Comparing our findings, which are in agreement with all other $z<1$
$f_{\mathrm{esc}}$ measurements, with $z\sim3$ studies
\citep{2006ApJ...651..688S,2009ApJ...692.1287I} implies that the
average escape fraction evolves with redshift, but the cause of this evolution remains unknown. It should be noted, however, that foreground contamination which is likely more severe at higher redshift, may account for some of the apparent evolution seen between $z=1$ and 3, by this work and others \citep{2010arXiv1001.3412S,2010MNRAS.404.1672V,2006MNRAS.371L...1I}.  
We proceed by focusing on possible mechanisms that could explain the lack of large escape fraction galaxies at $z\sim1$. 

\subsection{Selecting Analogs of High $f_{\mathrm{esc}}$ LBGs}
\label{sec:select-anal-high}

When comparing high- and low-redshift galaxy samples, there is always some degree
of uncertainty regarding the true analog nature of the two populations.   As discussed in Section \ref{sec:selection} and shown
in Figure \ref{fig:surf_bright}, we selected
a sample of LBG analogs sharing many of the same properties of the \citet{2006ApJ...651..688S} and \citet{2009ApJ...692.1287I} $z\sim3$ LBGs.
Figure \ref{fig:ebv_mass} further highlights their similarities showing the distribution
of reddening, stellar mass and rest-frame UV luminosity in these
sources is similar to the distribution in LBGs.  The similarity in
mass together with the UV--optical colors of the UVLGs implies that they may still be undergoing an early, major
episode of star formation rather than a small burst on top of a
hidden older population \citep[also see][for the same reason applied
to LBGs]{2004ApJS..154...97B}.  

Ultimately, this sample of LBG analogs
shares numerous similarities to the parent LBG population, but since only $\sim10\%-15\%$ of LBGs have been observed with
significant LyC detections, perhaps this subclass of LBGs has other
processes at work allowing or aiding in the escape of LyC photons.

\subsection{Galaxy Mergers}
\label{sec:galaxy-mergers}

Galaxy mergers offer an intriguing explanation for the increased
escape fraction seen at $z\sim3$.  \citet{2008ApJ...677...37O} noted that UVLGs
typically exhibit faint tidal features suggestive of a merger or recent
interaction.  They therefore propose that the super starbursts in
LBGs are triggered by gas-rich mergers.  Similarly, \citet{2009AJ....138..362P} showed that 20\%-30\% of LBGs have structures akin to local starburst mergers and may be driven by similar processes.  

As galaxies collide, strong gravitational and tidal forces can expel
long streams of stars, and ignite violent starbursts at rates of a few
to hundreds of $M_{\odot}$ yr$^{-1}$ \citep{1982ApJ...252..455S,2000ApJ...530..660B,2010ApJ...709.1067B}.
During a merger, the tidal fields distort the galaxies radially,
drawing out galactic material into long tails, plumes and bridges
\citep[e.g.,][]{1972ApJ...178..623T,1996ApJ...464..641M}.  The H I reservoirs can become disturbed, and the neutral
gas pulled away from the sources of ionizing radiation  producing
low-column density lines of sight \citep{2000AJ....119.1130H,2008AJ....135..548D} through the
galaxies, in turn allowing the escape of LyC photons.  Simulations by \citet{2008ApJ...672..765G} suggest that the escape
fraction in major mergers can be large ($f_{\mathrm{esc,rel}}\gsim30\%$)
compared to non-mergers ($f_{\mathrm{esc,rel}}<10\%$) along specific lines of sight.  Within our sample of 32
galaxies, 11 had morphologies consistent with merger activity.  We
independently stacked eight of these spectra (removing three due to their larger spatial extent) and find
$f_{\mathrm{esc,rel}}<2\%$ ($3\sigma$ upper limit).  

If merging is a viable mechanism for clearing pathways in the
ISM for LyC photons, the orientation of the system along the line of
sight is also a likely factor, requiring 
a large sample of UV luminous mergers.   Therefore, we cannot say
whether mergers are an important factor as our sample size is
at present too small.  Currently, there is a lack of deep high-resolution rest-frame optical imaging of the LBGs with larger escape
fractions, and the interpretation of UV morphologies remains
problematic \citep{2007ApJ...656....1L}. Future near-IR observations with \hst\ of the $z\sim3$ LBG leakers will shed light on this hypothesis.

As discussed earlier in the section, galaxy mergers are capable of
clearing pathways, exposing UV bright stars.  If mergers do facilitate the
escape of LyC radiation then an evolving merger rate, may be
responsible for the observed evolution in $f_{\mathrm {sc,rel}}$.  Numerous
observational studies and simulations haven shown that the galaxy
merger rate evolves with redshift, going as $\sim(1+z)^{2-3}$
\citep{2001ApJ...546..223G,2003AJ....126.1183C,2007ApJS..172..320K,2007ApJ...659..976H,2007ApJ...659..931B,2008MNRAS.386..909C,2010ApJ...709.1067B}.
 The factor of 3-4 increase in merger rate between $z\sim$1 and 3
 as seen observationally,  would also increase the number of lines of sights and
 range of encounter parameters observed in $z\sim3$ galaxy mergers by
 the same factor.   This would in turn  increase the likelihood of
 detecting LyC at higher redshift.

\subsection{Evolving $f_{\mathrm{esc,rel}}$: Size, Mass, and Star Formation}
\label{sec:an-evolving-escape}

The LyC escape fraction is limited by the distribution of neutral hydrogen along a line of sight and likely depends on galactic parameters.
We now consider what galaxy
properties could evolve with redshift that reduce the efficiency of
galactic outflows/chimneys in leaking LyC
photons from luminous galaxies.

Typical galaxies (including UV bright galaxies) have been shown to be 1.5-3 times smaller at $z\sim3$ than their local
counterparts \citep{2006ApJ...650...18T,2002ApJ...579L...1P,2004ApJ...600L.107F}. 
Although our sample was selected to have similar UV
surface brightnesses as $z\sim3$ LBGs (refer to Figure
\ref{fig:surf_bright}), little is known about the true optical sizes
of LBGs or the gas distribution.  The velocities of galactic winds or
outflows have been found to be proportional to the SFR in LBGs
\citep{2006MNRAS.373..571F}, therefore LBG and LBG analogs, having
similar SFR should in principle generate outflows with similar
velocities (a few hundred km s$^{-1}$).  However, smaller galaxies would have
higher SFRs per unit volume,  which can result in more
efficient galactic winds \citep{2005ARA&A..43..769V}, more easily
clearing pathways or ``chimney-like'' structures,  and in turn
allowing for higher $f_{\mathrm{esc,rel}}$ \citep{2003ApJ...599...50F}. 

With smaller galaxies, and higher-density starbursts comes the potential for a
larger fraction of stars born in very compact star clusters, including
super star clusters (SSCs).  SSCs can have thousands of young ($<$50Myr)
stars within a half-light radius of $\sim$10pc \citep{2005ARA&A..43..769V}.  These extreme
concentrations of hot O and B stars can greatly impact the state of
the ISM driving powerful galactic winds (like those seen in M82), opening channels
for LyC photons to escape.   SSCs have been detected in the tidal
tails \citep{2009AAS...21334401C} and outer regions of galaxies, which could explain the spatially offset LyC emission (to the
optical emission) detected by \citet{2009ApJ...692.1287I} in a few
$z\sim3$ LBGs.    There is also some evidence that SSCs found in the
local group of galaxies have top-heavy initial mass functions
\citep[IMFs;][]{2008ApJ...675.1319H}, which enhance the efficiency in clearing
lines of sight, due to the larger outflow velocities generated by massive stars.   

Another property which has been shown to possibly evolve with redshift  is the
stellar mass-to-dark matter ratio.  Recent work by
\citet{2010MNRAS.401.2113E}, have reported that local
massive galaxies have more centrally concentrated dark matter than higher redshift galaxies
($z\sim0.7$) with comparable stellar masses.  A higher stellar mass-to-dark matter mass ratio at lower redshift would result in a larger gravitational potential, 
and in turn the requirement for larger galactic winds to achieve the same ``porosity'' of the ISM (reducing the probability for ionizing photons to escape their host galaxies).  
Thicker disks due to dense concentrations of gas have also been shown in to impede or slow the development of galactic outflows, due to a higher gravitational potential \citep{2008ApJ...674..157C}.

\subsection{Intrinsic LyC-to-UV Ratio: A Top-heavy IMF?}
\label{sec:evolv-f_esc:-intr}

An evolving intrinsic LyC-to-UV flux density ratio could also be
responsible for the observed evolution in $f_{\mathrm{esc,rel}}$.  The escape
fraction is measured by comparing the UV (1500\AA) flux of a galaxy
to flux below the Lyman limit.  One key assumption that is made
involves the inherent LyC-to-UV ratio of starbursts.  Typically, this ratio is
considered constant with redshift, but if there were an order of
magnitude larger production rate of the LyC relative to the 1500\AA, flux
at $z>3$ than at $z<1$, the observed change in $f_{\mathrm{esc,rel}}$ would be expected. 

\citet{2009ApJ...692.1287I} found that three of the six LBGs with detected escaping LyC in their sample,
had SEDs intrinsically bluer than those
expected from population synthesis models, assuming a standard IMF
with moderate dust attenuation.  They show that an intrinsically bluer
SED ($\sim0.3$ mag bluer in NB359-R than those of
Starburst99), in the absence of QSO activity, can be produced with a top-heavy IMF.
They also suggest that a deviation  from the
\citet{2000ApJ...533..682C} dust attenuation law, with less dust
absorption at 900\AA, compared to that at 1500\AA, can come close to
achieving the observed bluer colors of those LBGs.  

It has been shown that the UV luminosity density
at $z\sim6$ may be insufficient to explain the ionized universe at
$z>7$ unless the IMF allowed for the production of more massive
stars \citep{2008ApJ...680...32C}.  A top-heavy IMF at higher redshift could help explain the
larger number of LyC detections at $z\sim3$ as massive stars produce
more LyC photons, and stronger supernova driven winds.  A top-heavy
IMF requires regions of potentially low metallicity gas which is
consistent with the metallicity and dust evolution seen from $z\sim3$
to 1 \citep{2002ApJ...569L..65F}.

\section{Summary}
\label{sec:summary}

LyC photons produced in massive starbursts likely played an important
role in the reionization of the universe.  However, their
contribution depends upon the fraction of ionizing radiation that can escape the high column density of HI gas surrounding these star-forming galaxies.  We have presented \hst\ rest-frame UV slitless spectroscopy of 32 $z\sim$0.7 LBG analogs
in the COSMOS field to investigate the LyC escape fraction.  These UV spectra have achieved the deepest
limits to date on the escape fraction in individual sources at any
redshift.  A summary of our results is as follows.

Aside from the detection of leaking LyC from an AGN--starburst composite, we find no detections of LyC in our sample of 31 star-forming galaxies. 
The individual $3\sigma$
lower limits of $f_{\nu}$(1500\AA\,)/$f_{\nu}$(830\AA\,) ratio range from 20 to
204 (median of 73.5) and 378.7 in the stack of 18 galaxies.  Assuming
an intrinsic Lyman break of 3.4 and an IGM transmission of $85\%$, we
report a {\it relative} escape fraction in individual
galaxies of 0.02-0.19 and 0.01 in the stack ($3\sigma$ upper limit).
There is no indication of the near unity escape fractions found at
$z\sim3$.  The striking contrast between the nearly zero escape
fractions found in the 22 $z\sim0.7$ LBG analogs with the near unity escape
fractions discovered in 10\% of the $z\sim3$ LBG population
strongly argues for an evolving escape fraction.   It is unclear,
however, if the lack of near unity escape fraction detections at low
redshift is due to an evolution in actual value of the $f_{\mathrm{esc,rel}}$ itself, 
or if it is just that the number of galaxies that in fact have large amounts of leaking LyC, decrease with redshift.  Both scenarios could explain absence of galaxies with larger amounts of leaking LyC.    Possible causes for
a change in the perceived escape fraction with redshift involve a top-heavy IMF, larger SFR densities,
stellar mass-to-dark matter ratios, 
and/or fraction of SSCs at higher
redshifts.  All these mechanisms enhance the efficiency of galaxy winds, increasing
the porosity of the ISM, facilitating LyC escape.   However,
if galaxy mergers aid in the escape of LyC radiation then an evolving galaxy merger rate could account for the high number of LyC leaking galaxies at $z\sim3$.  The lack of low-redshift galaxies with escaping LyC could then be explained by the small number of galaxy mergers that have been observed below the Lyman limit.  An additional consideration is that a fraction of the detections of escaping LyC at $z\sim$3 may be a consequence of foreground contamination, which would reduce the strength of the evolution in $f_{esc,rel}$ seen when comparing to low-$z$ studies.

The escape fraction of UV radiation in $z\sim1$ luminous starburst galaxies is
an important quantity to understand since it provides insight into the
sources (massive star formation or QSOs) responsible for reionization.
Our study has presented a robust measure of the
$f_{\mathrm{esc,rel}}$ in the low-redshift universe and suggests that the escape
fraction in objects that are analogous to $z\sim$3 LBG population, evolves with redshift.  However, future
study is required to isolate the cause of this evolution.

\acknowledgments
We thank Mark Dickinson and Colin Borys for their contributions to this work and the anonymous referee for their constructive comments which added to the clarity of this paper.
The research described in this paper was carried out, in part, by the
Jet Propulsion Laboratory, California Institute of Technology, and observations obtained at the Hale Telescope, 
Palomar Observatory as part of a continuing collaboration between the California Institute of Technology, NASA /JPL, and 
Cornell University. 
Support for programs HST-GO 11236 was provided by NASA through
grants from the Space Telescope Science Institute, which is operated
by the Association of Universities for Research in Astronomy, Inc.


\bibliographystyle{apj}
\bibliography{astro}

\clearpage

\begin{figure}[h]
\begin{center}
\caption[Spectra of $z\sim0.7$ LBG Analogs]{Deep UV slitless spectra
  of $z\sim0.7$ LBG analogs in the COSMOS field.  The spectra have
  been shifted into the rest frame and fluxes are in
  erg cm$^{-2}$ s$^{-1}$\AA$^{-1}$. The horizontal lines (red) represent the average flux
  over two wavelength ranges, $\sim$780-880\AA\ and 1000-1050\AA\ with 3$\sigma$ error bars.}

\vspace{0.5cm}
\includegraphics[width=150mm]{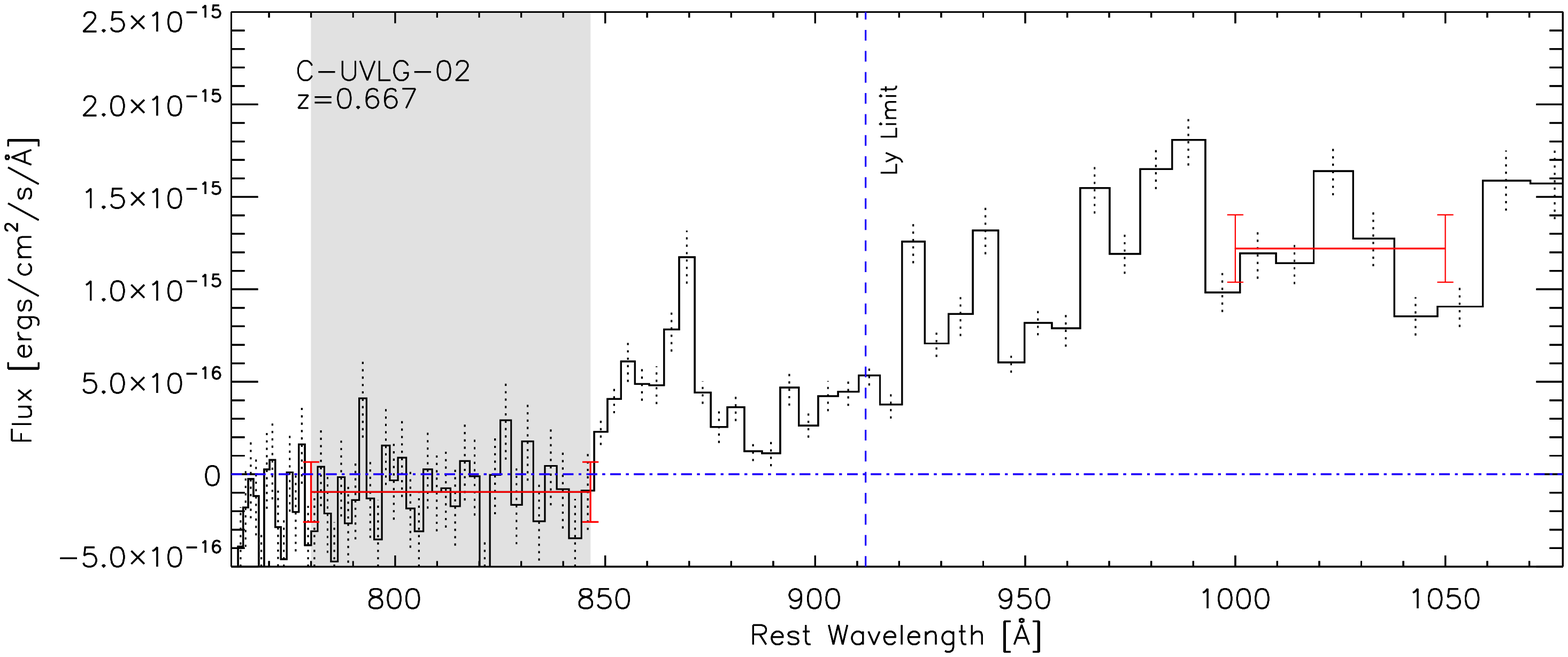}
\includegraphics[width=150mm]{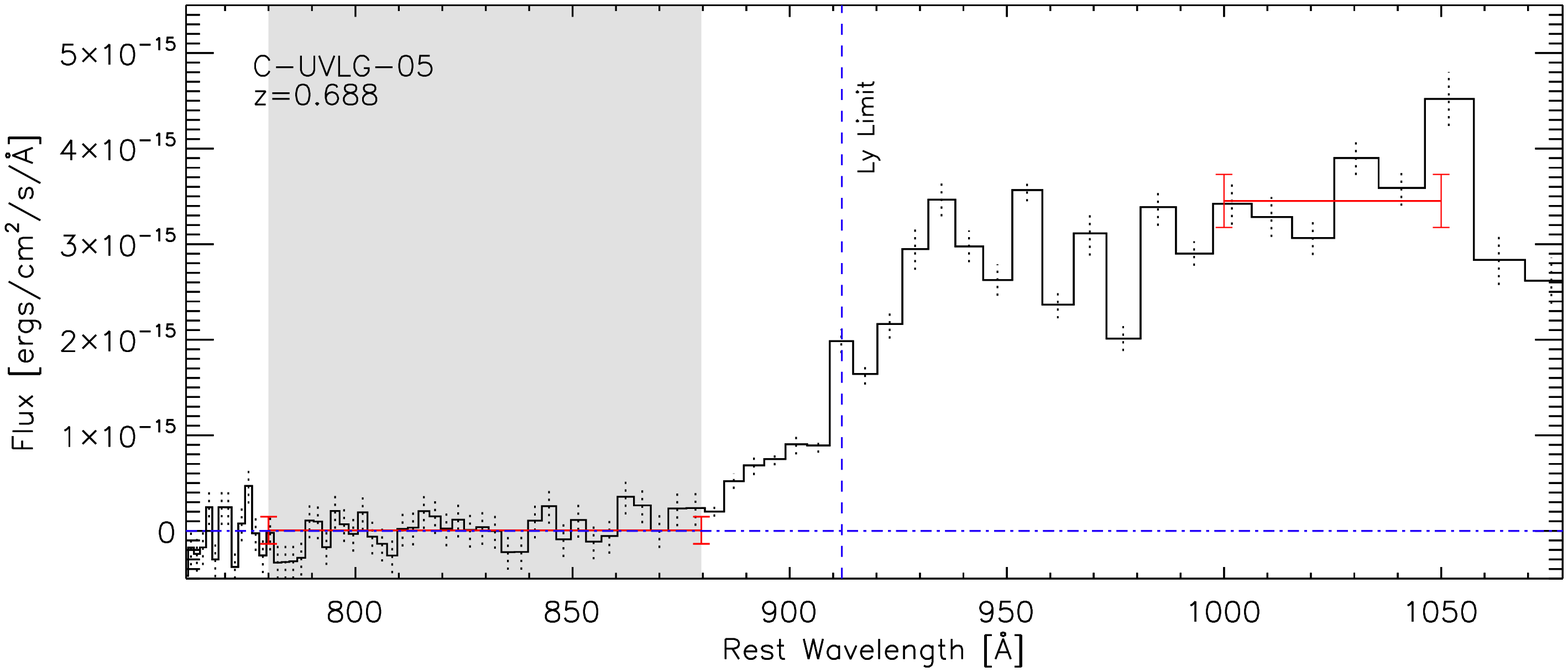}
\includegraphics[width=150mm]{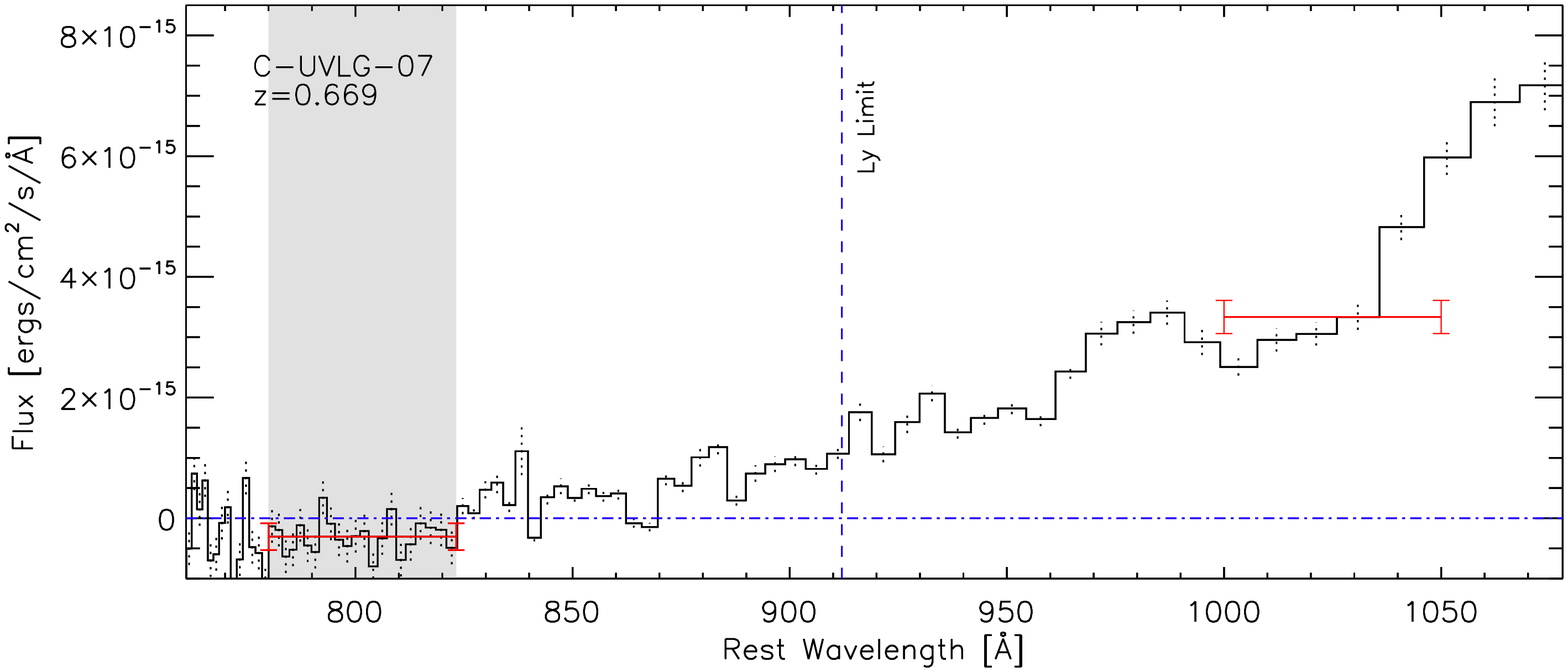}
\end{center}
\label{fig:fullsamplespectra}
\end{figure}

\begin{figure}
\begin{center}
\includegraphics[width=150mm]{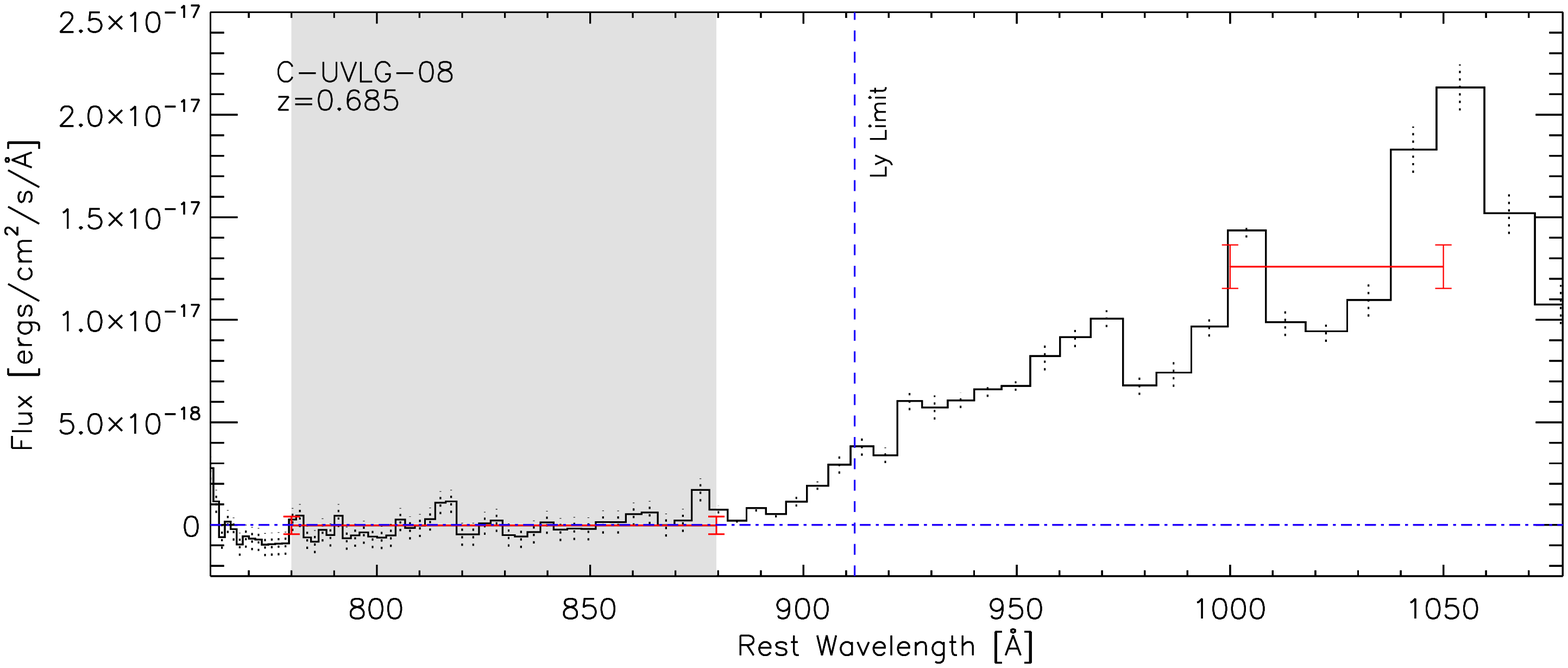}
\includegraphics[width=150mm]{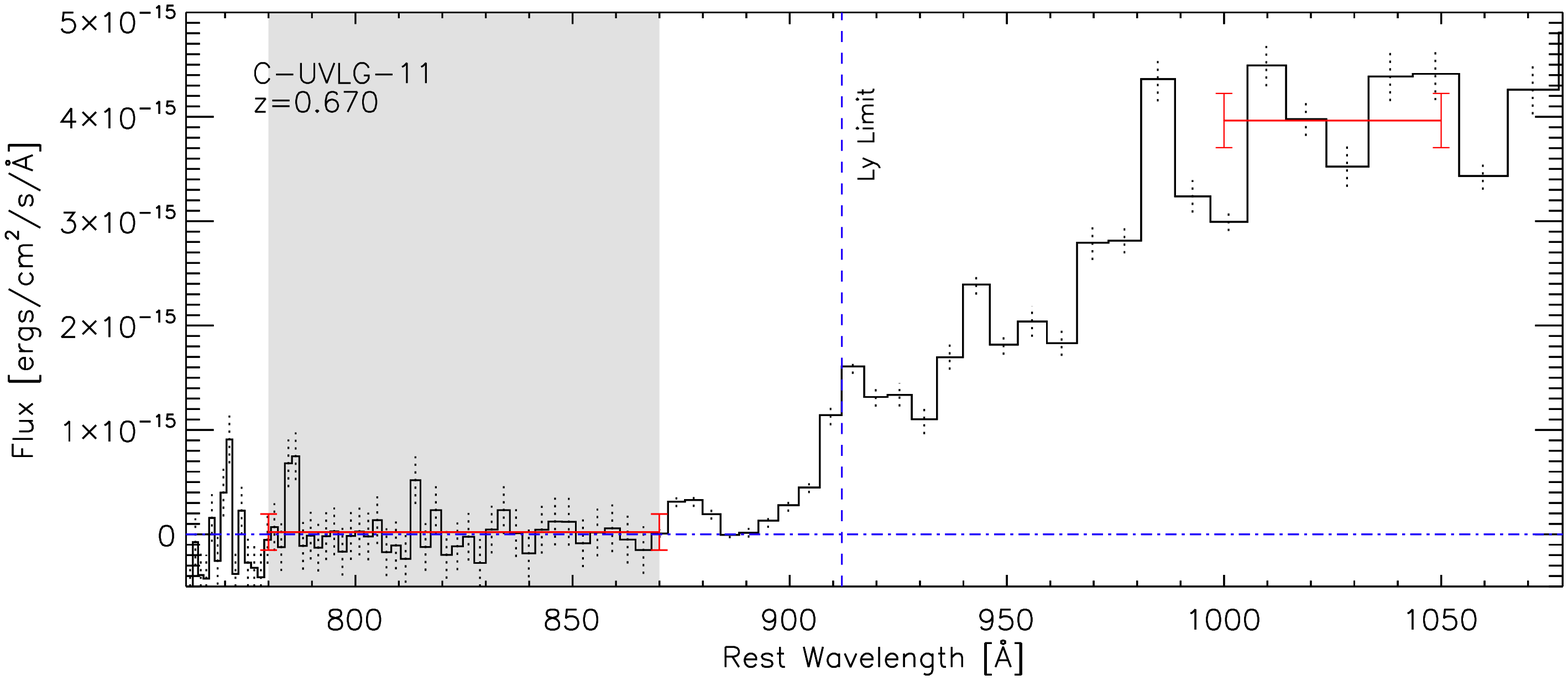}
\includegraphics[width=150mm]{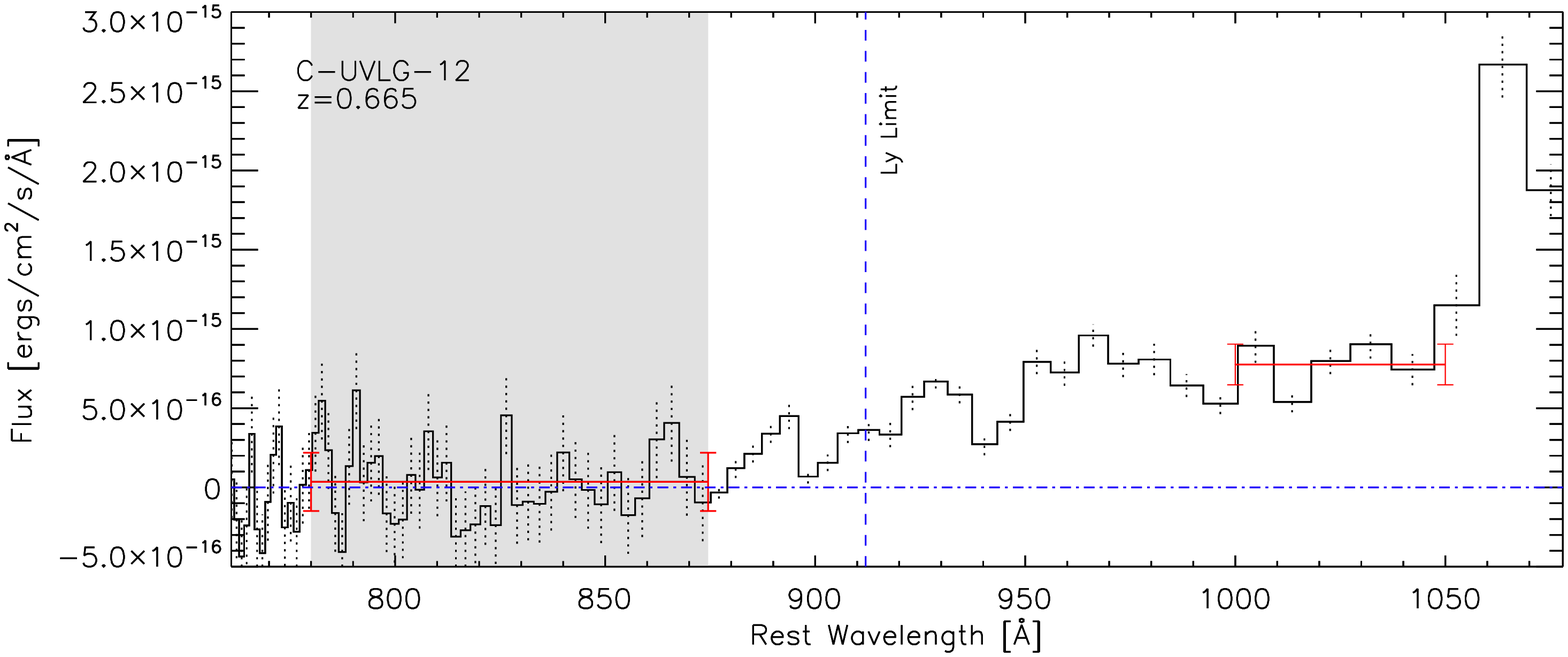}
\end{center}
\end{figure}

\begin{figure}
\begin{center}
\includegraphics[width=150mm]{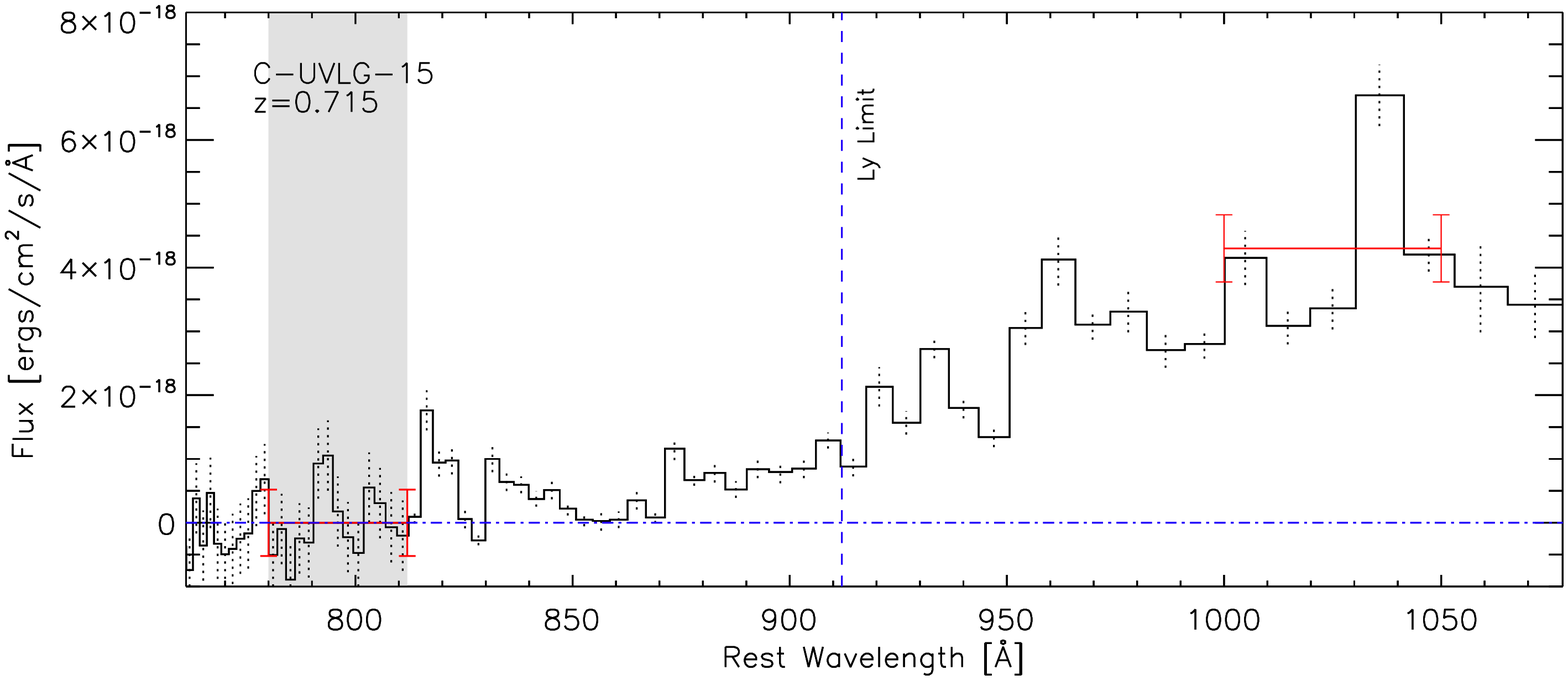}
\includegraphics[width=150mm]{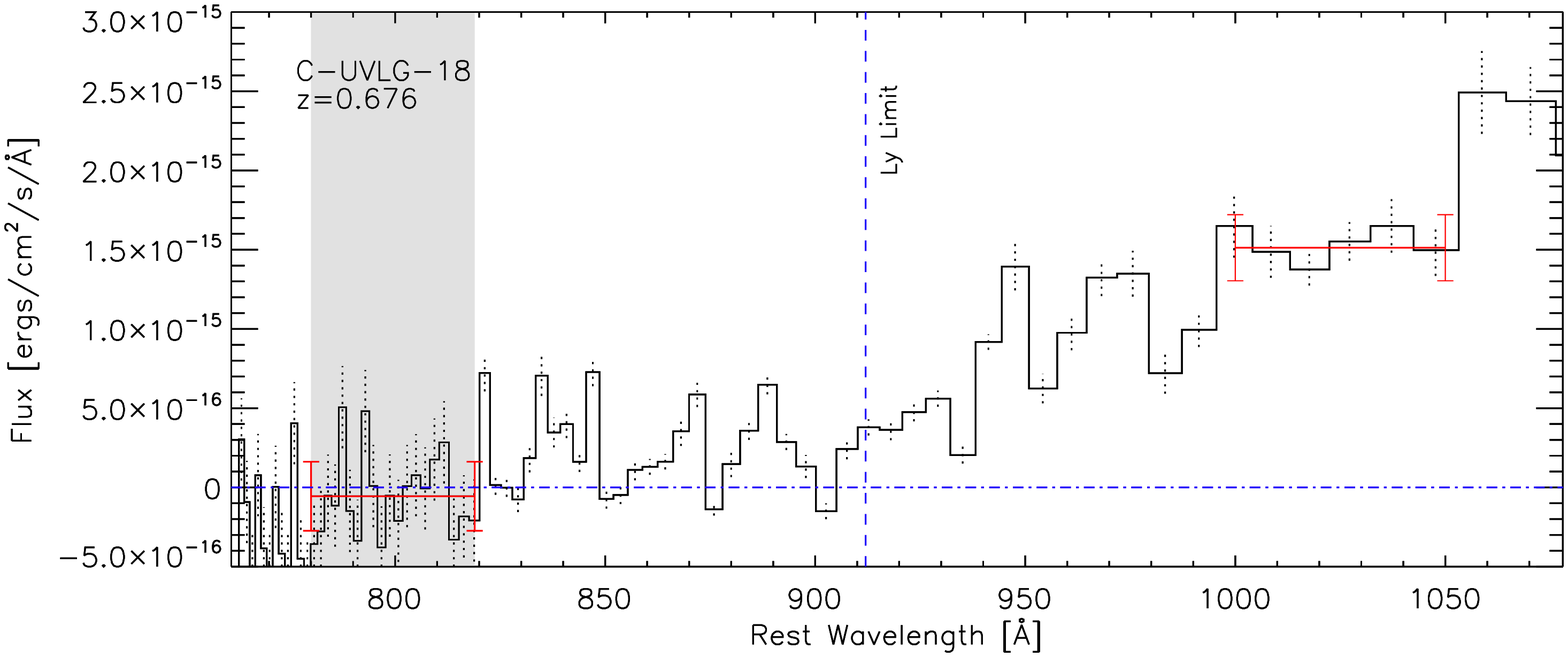}
\includegraphics[width=150mm]{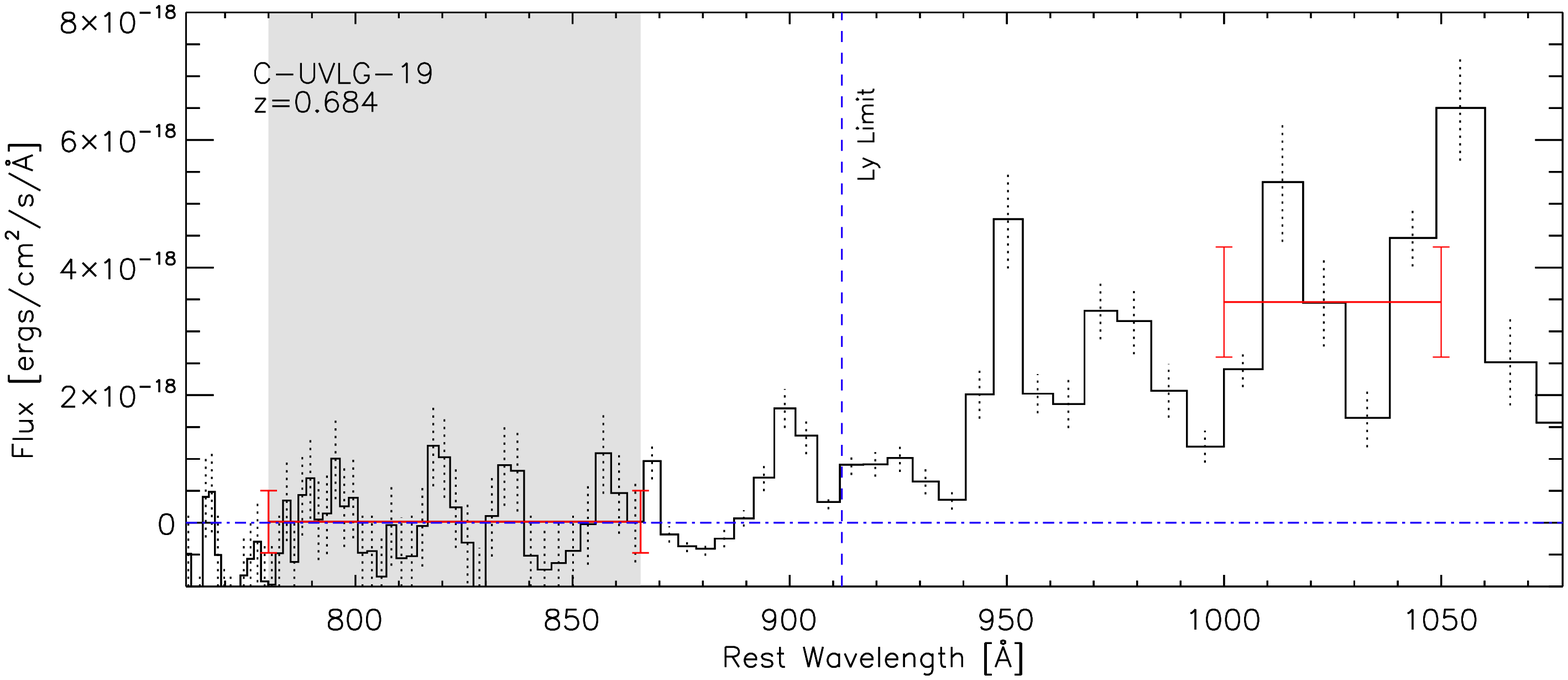}
\end{center}
\end{figure}

\begin{figure}
\begin{center}
\includegraphics[width=150mm]{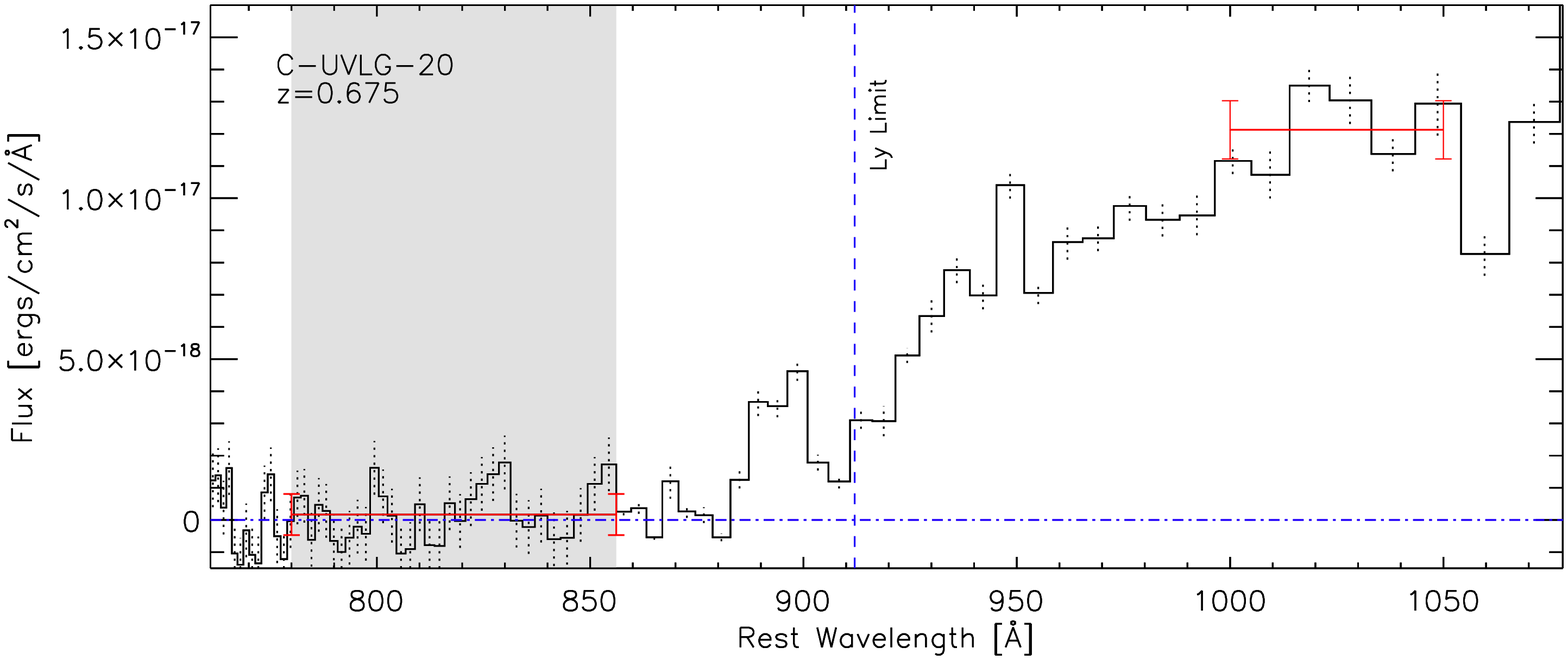}
\includegraphics[width=150mm]{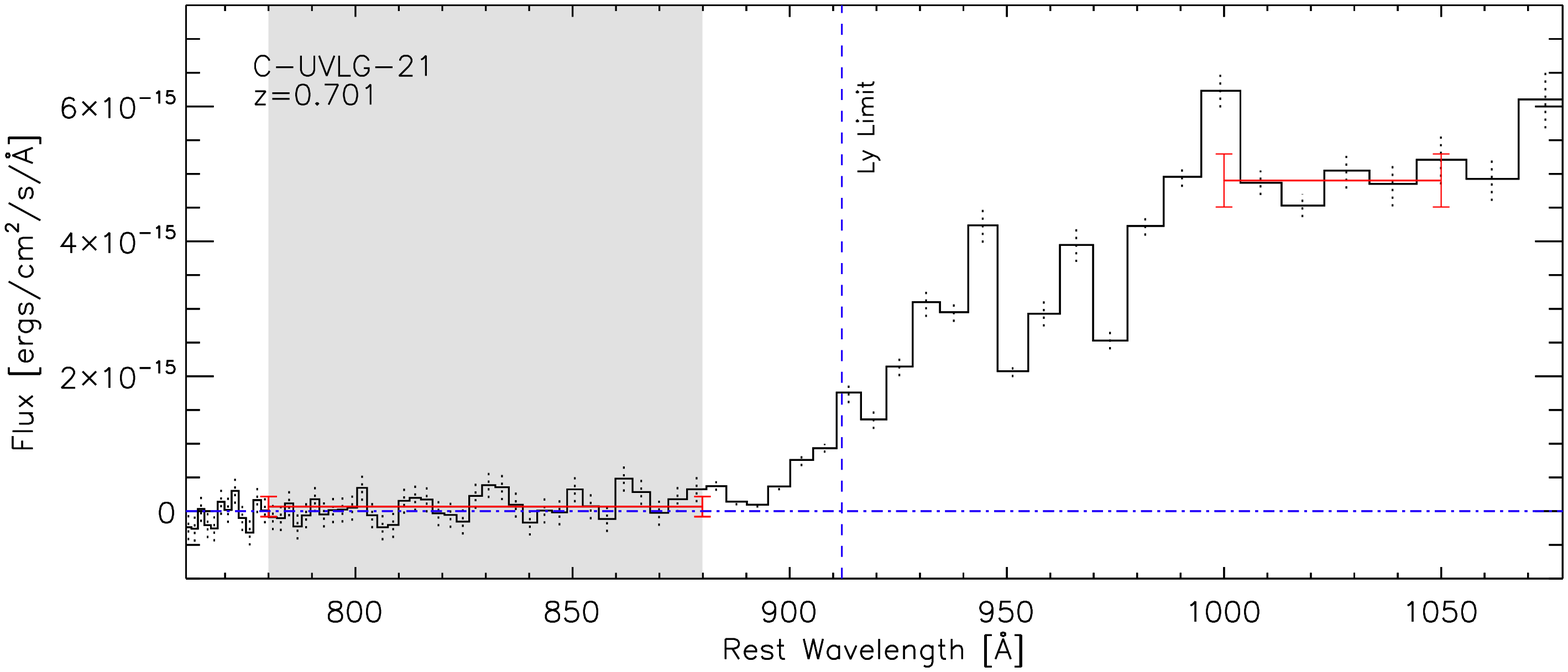}
\includegraphics[width=150mm]{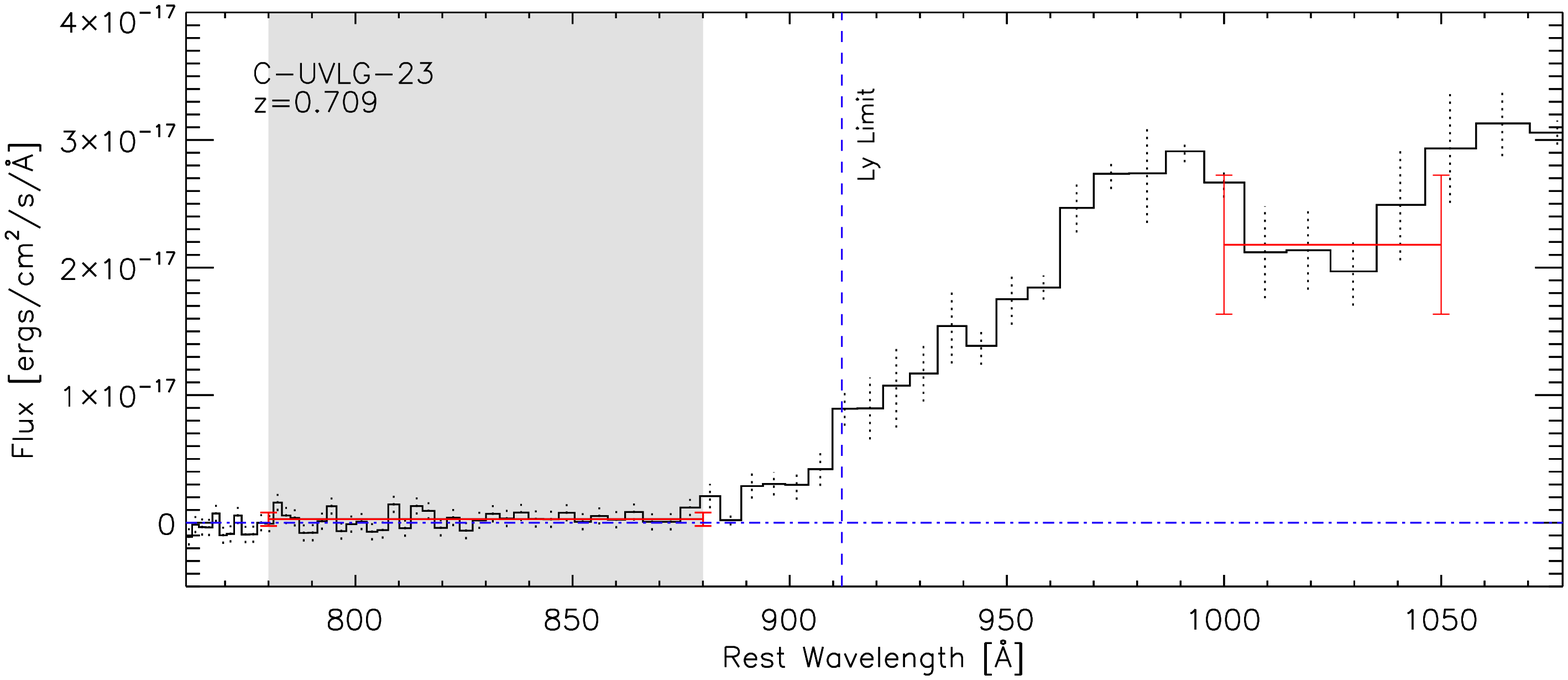}
\end{center}
\end{figure}

\begin{figure}
\begin{center}
\includegraphics[width=150mm]{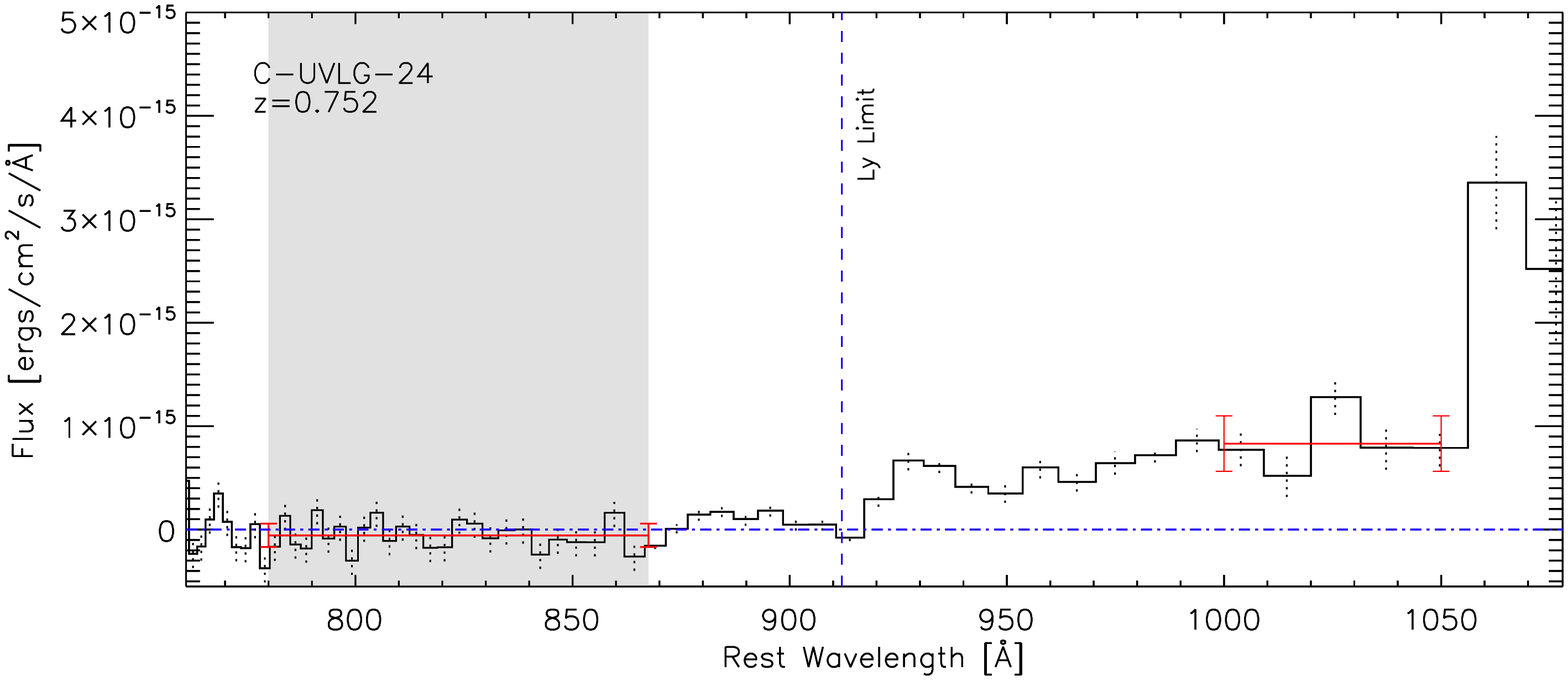}
\includegraphics[width=150mm]{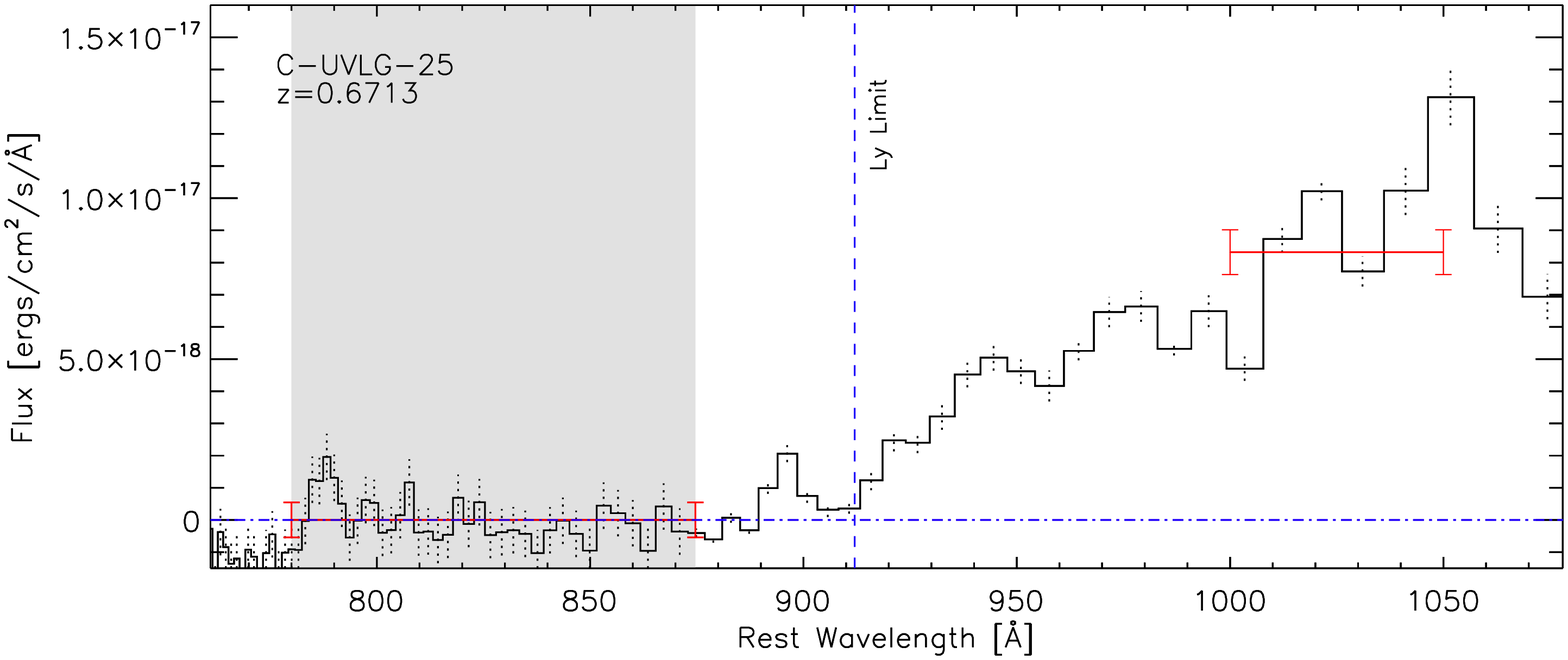}
\includegraphics[width=150mm]{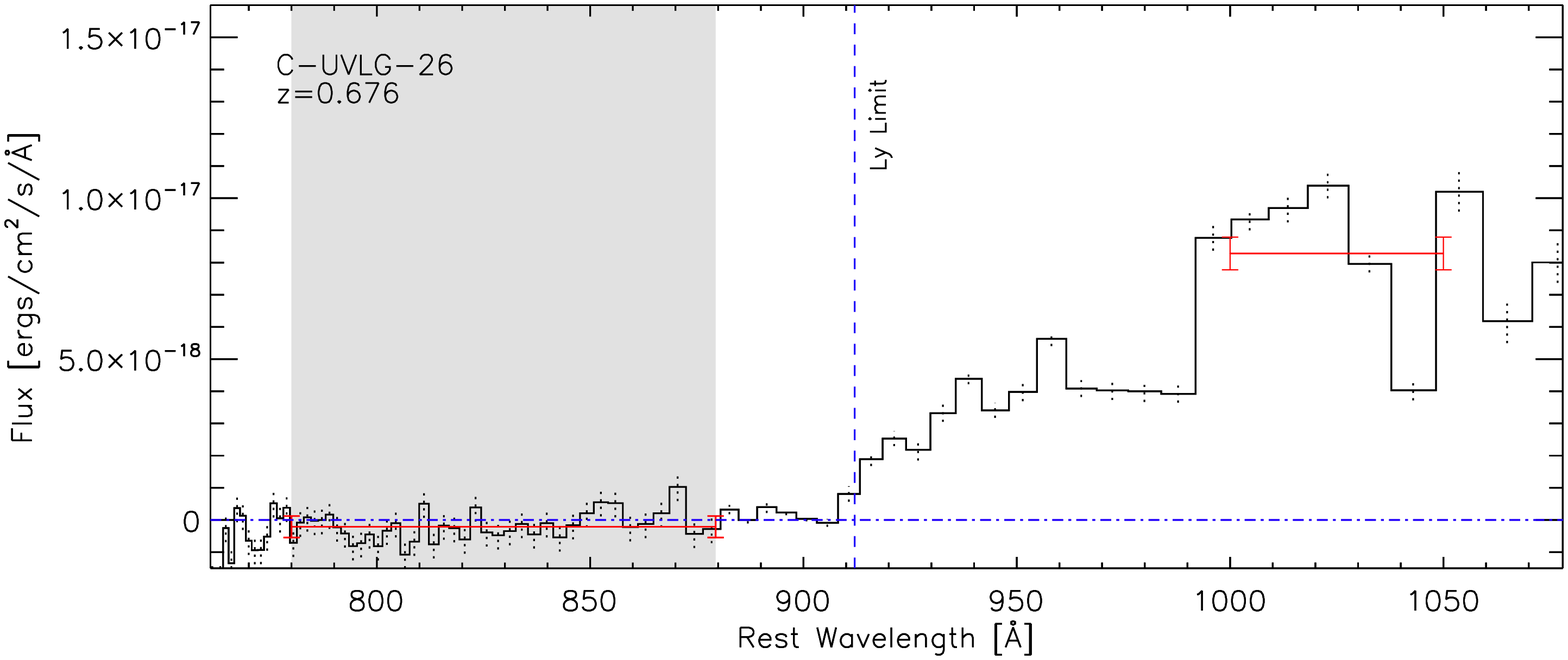}
\end{center}
\end{figure}

\begin{figure}
\begin{center}
\includegraphics[width=150mm]{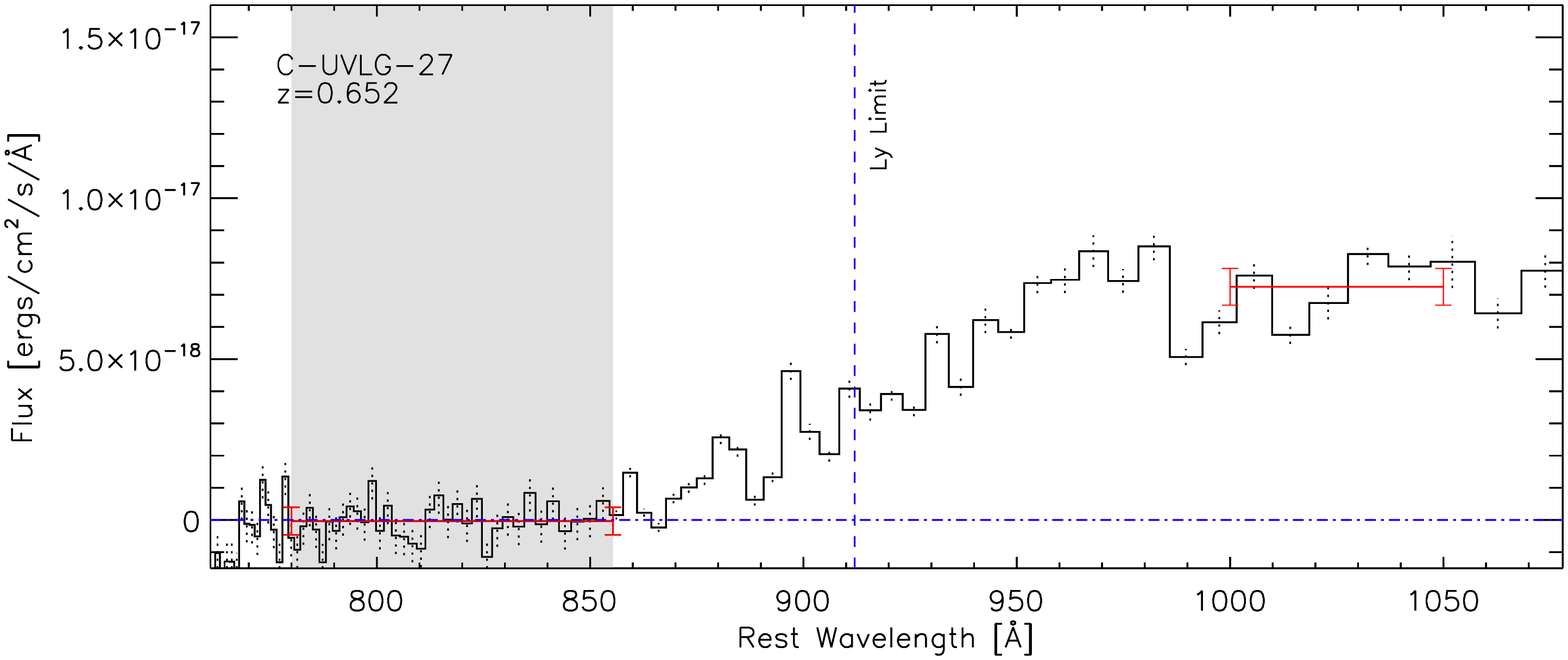}
\includegraphics[width=150mm]{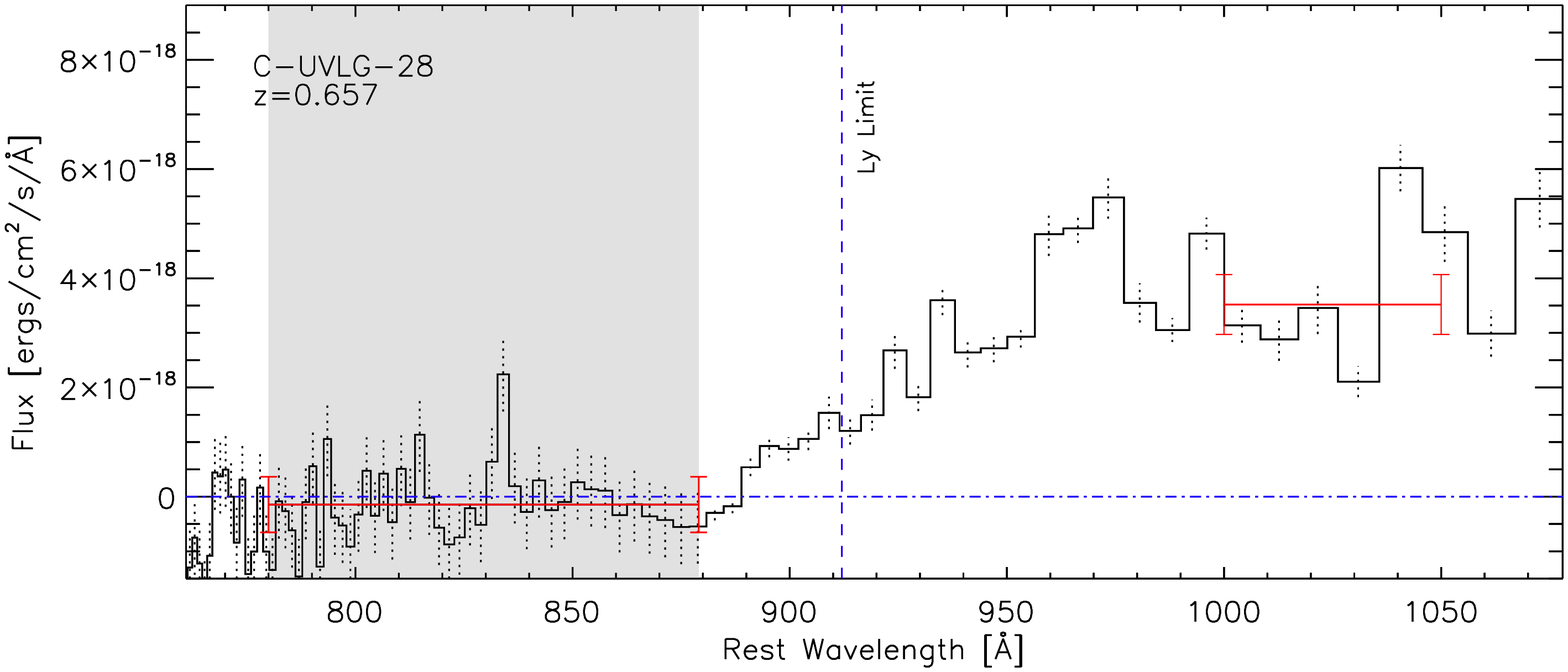}
\includegraphics[width=150mm]{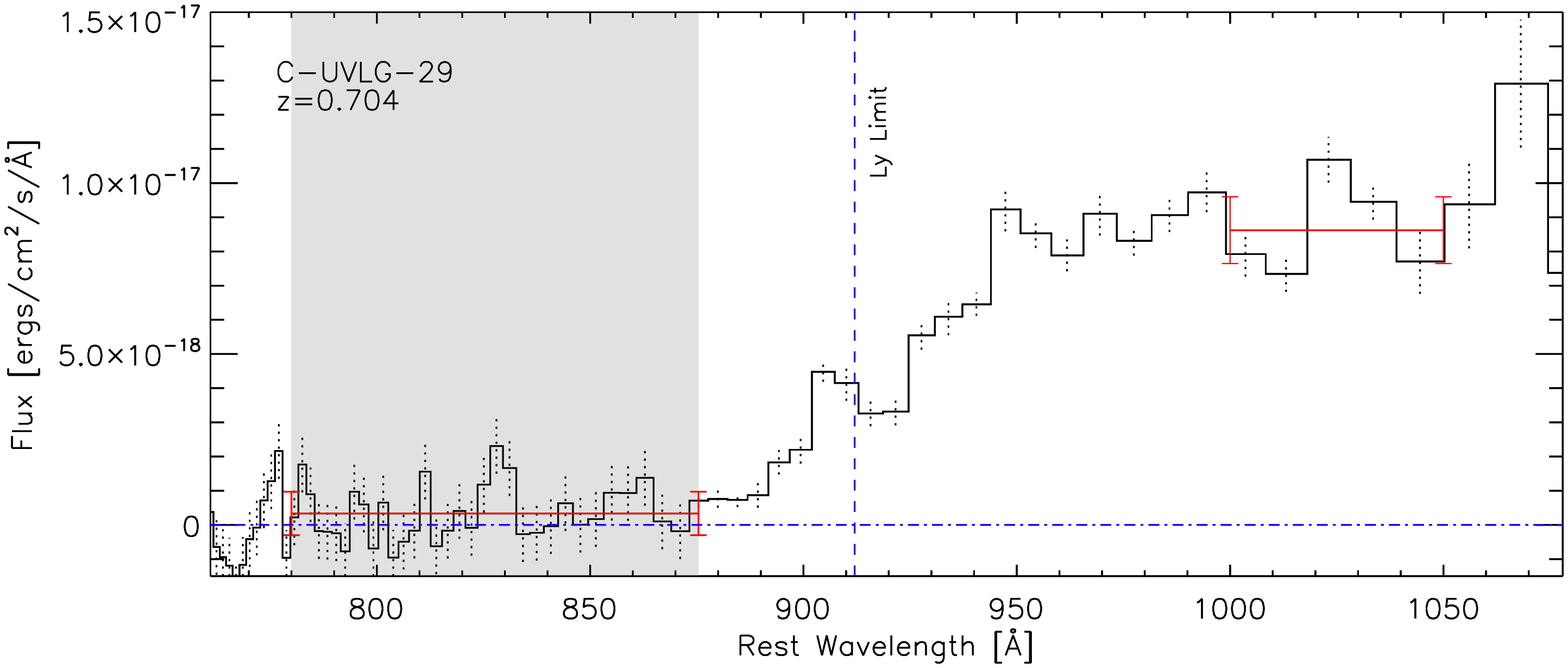}
\end{center}
\end{figure}

\begin{figure}
\begin{center}\hspace{0.1cm}

\includegraphics[width=150mm]{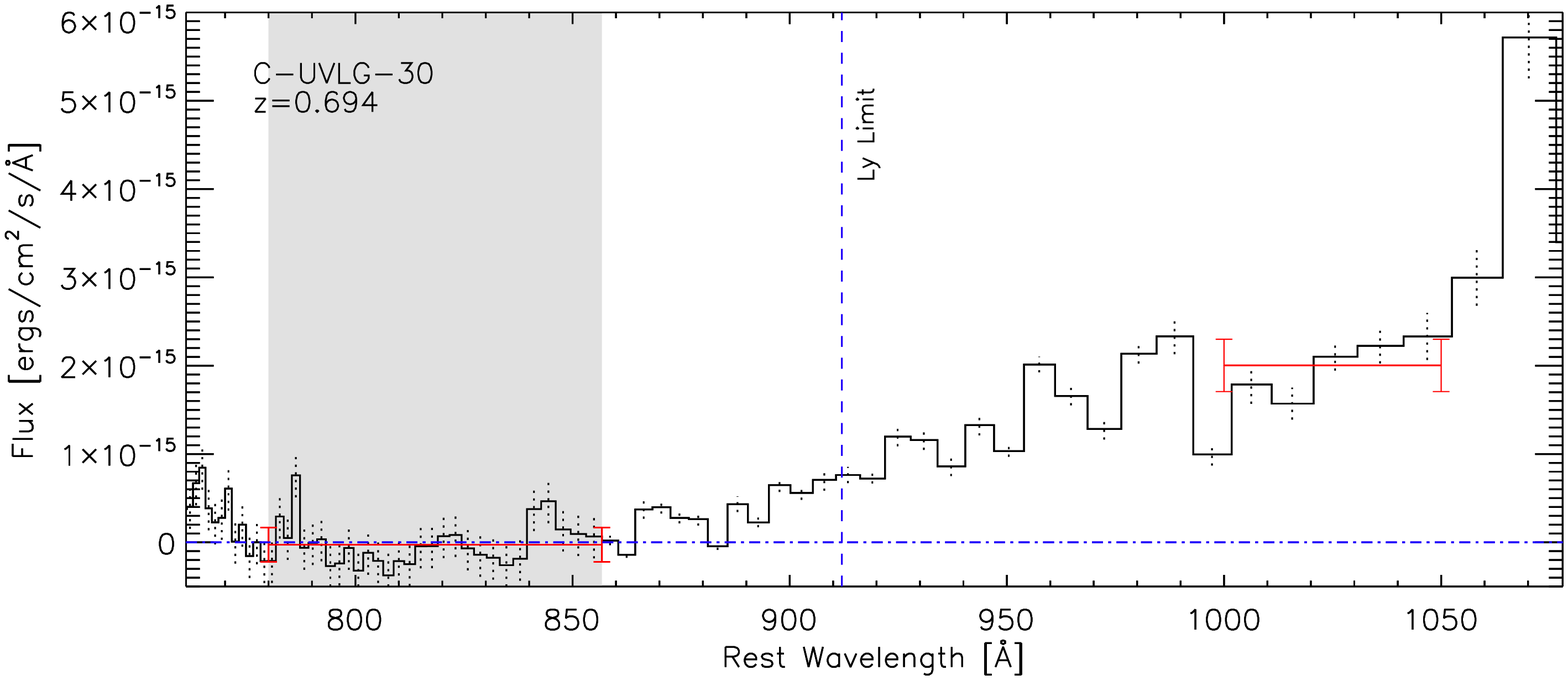}
\includegraphics[width=150mm]{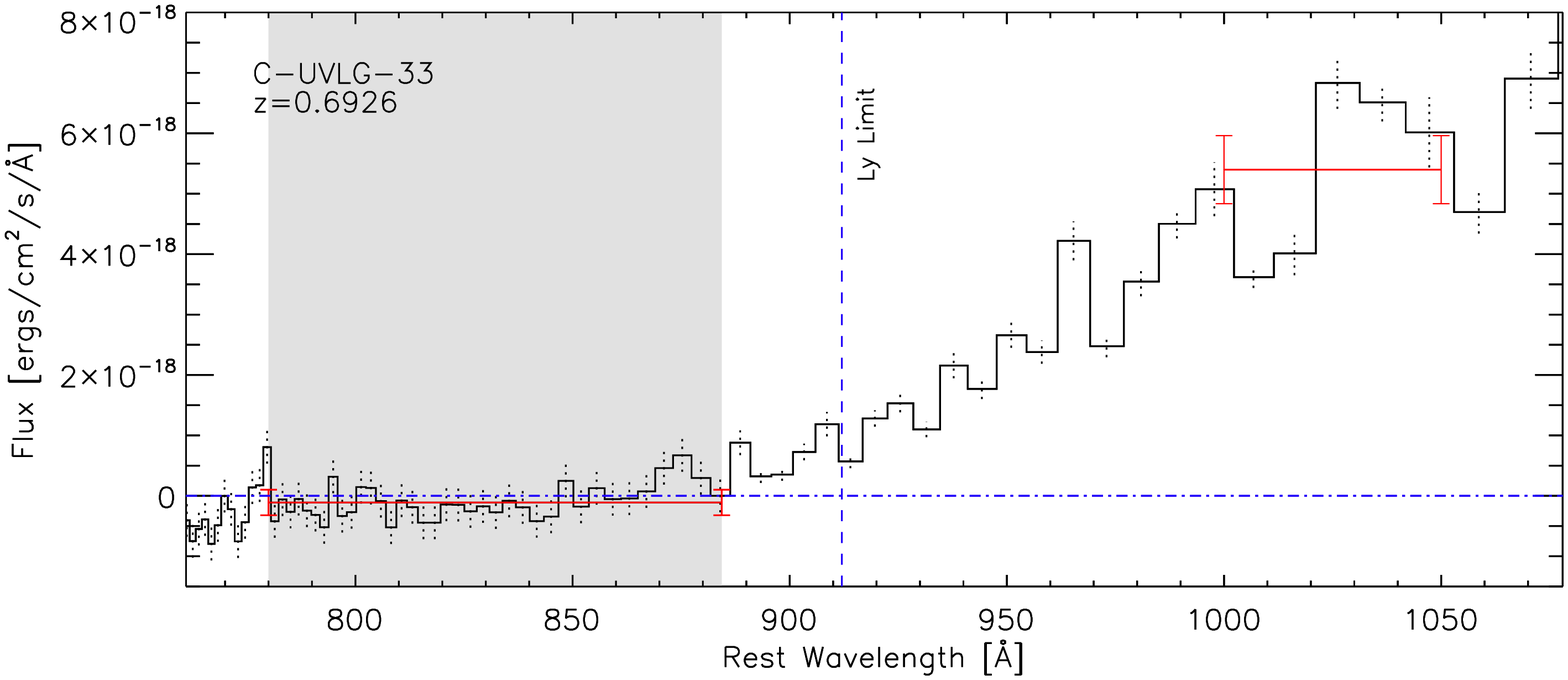}\\
\includegraphics[width=150mm]{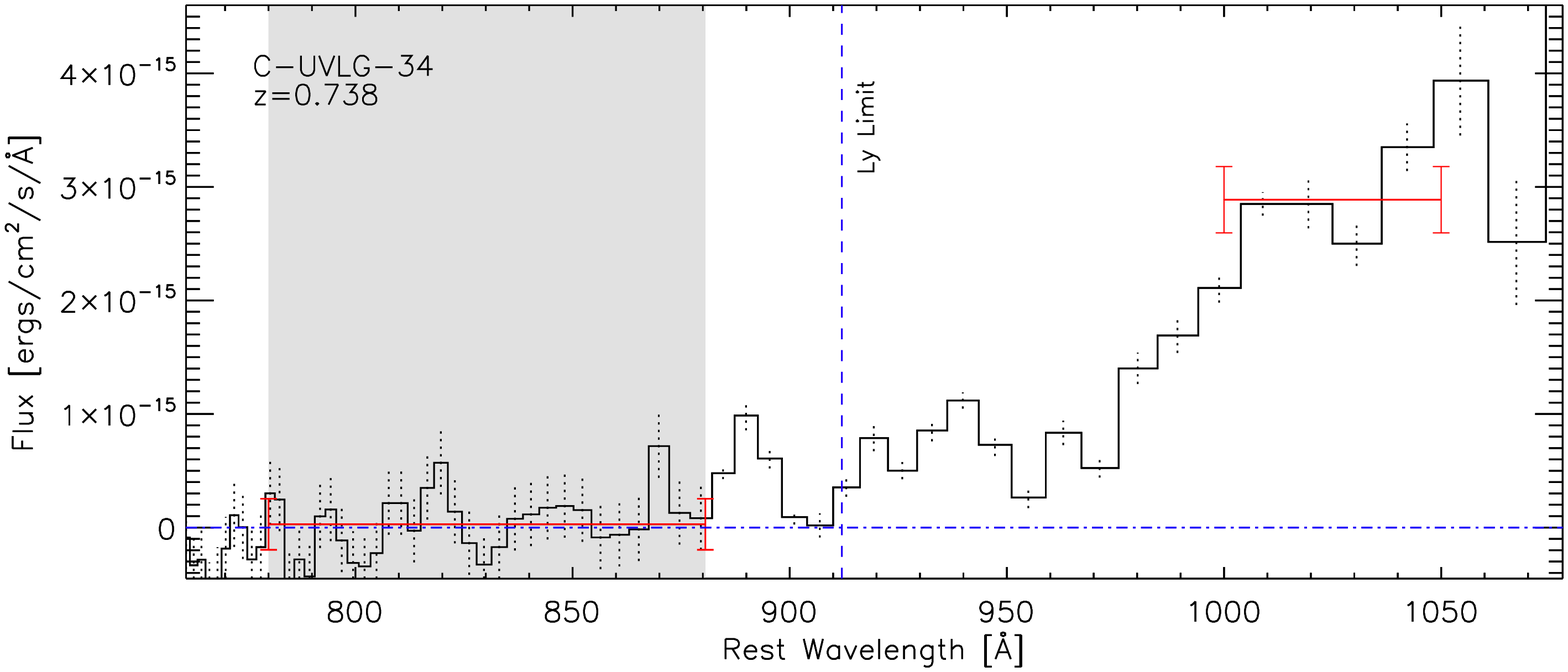}
\end{center}
\end{figure}

\begin{figure*}
\begin{center}
\includegraphics[width=150mm]{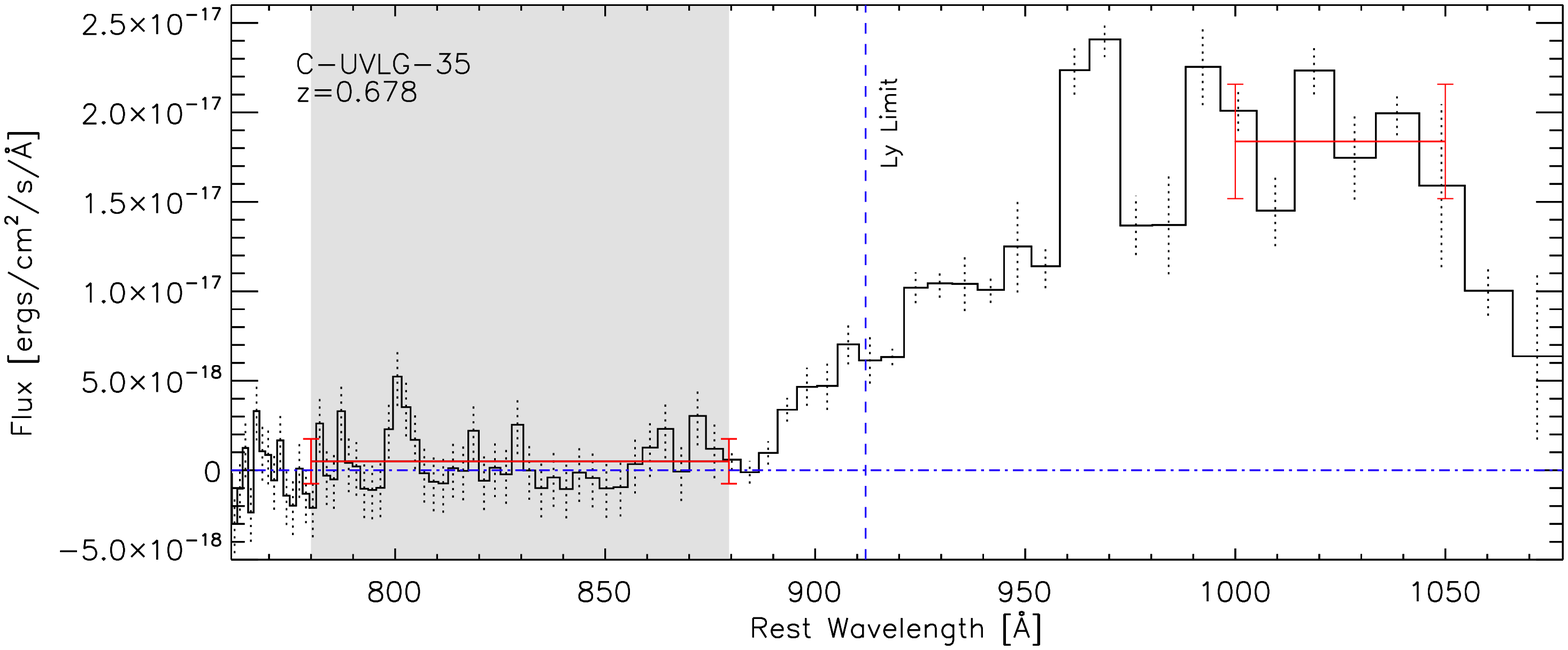}
\end{center}
\end{figure*}


\begin{figure}
  \centering
  \includegraphics[width=150mm]{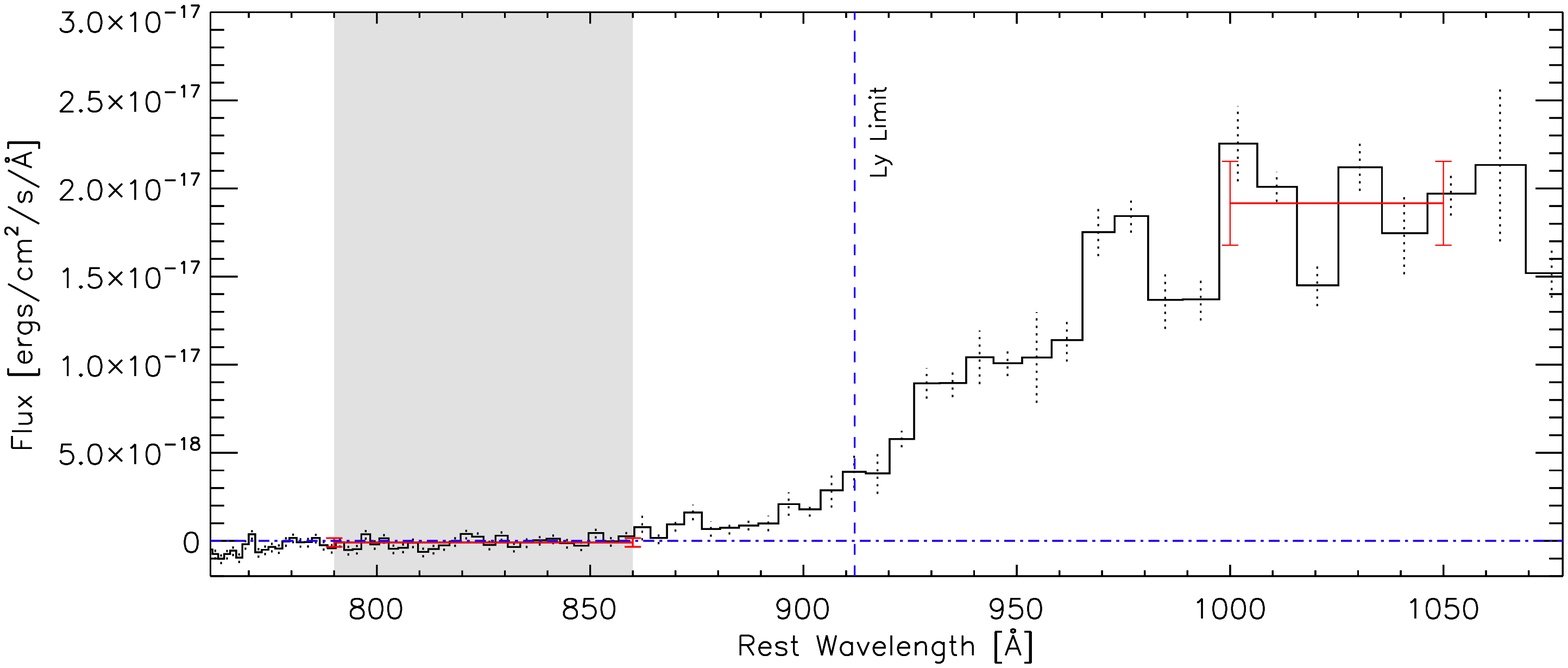}
\caption[Composite Spectrum]{Stacked spectrum of 18 $z\sim 0.7$
  LBG analogs.  The $3\sigma$ lower limits on the observed UV-to-LyC
  flux density ratio is 378.7.  Assuming an intrinsic Lyman break of
  3.4 the upper limit on the {\it relative} escape fraction is 0.01 ($3\sigma$).  The Lyman limit is indicated by the vertical dashed
  line.  The horizontal lines (red) represent the average LyC and $f_{1025}$ flux
 used to derive the escape fraction.  Error bars are discussed in Section \ref{uv spec} and are the same as in Figure 5.}
\label{fig:stackedfull}
\end{figure}

\begin{figure}
  \centering
  \includegraphics[width=150mm]{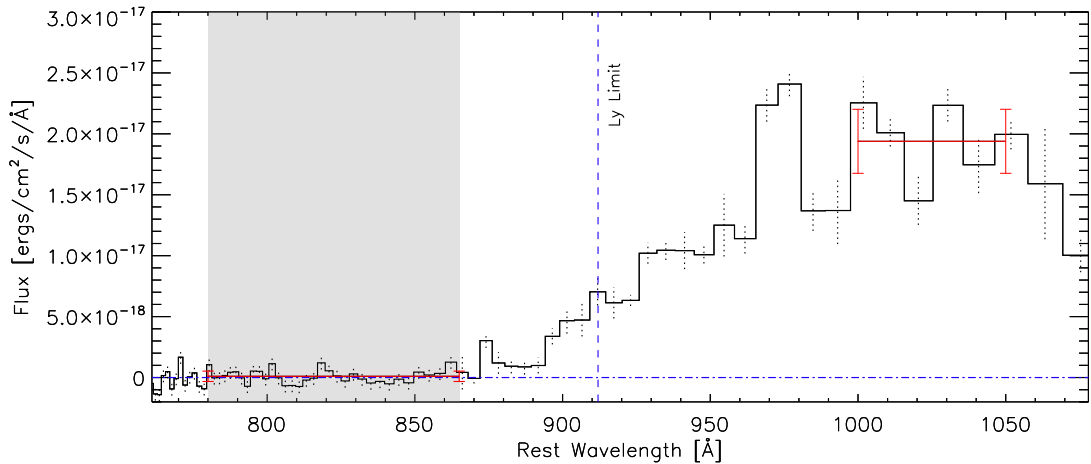}
\caption[Composite Spectrum]{Stacked spectrum of 8 $z\sim 0.7$
  LBG analogs with a merger morphology.  The $3\sigma$ lower limits on the observed UV-to-LyC
  flux density ratio is 223.2.  Assuming an intrinsic Lyman break of
  three the relative escape fraction is 0.02 ($3\sigma$).  The Lyman limit is indicated by the vertical dashed
  line.  The horizontal lines (red) represent the average flux
  over two wavelength ranges, 800-860\AA\ and 1000-1050\AA\ with 3$\sigma$ error bars.}
\label{fig:stackedmerger}
\end{figure}

\renewcommand{\thefootnote}{\alph{footnote}}
\scriptsize
\begin{center}
\begin{deluxetable}{lccccccccc}
\tabletypesize{\scriptsize}
\tablecaption{\label{tab:result}SBC Observations of LBG Analogs in the COSMOS Field}

\tablehead{\colhead{Object ID}& 
\colhead{R.A.} & 
\colhead{Decl.} & 
\colhead{Exp. Time}& 
\colhead{$z_{\mathrm{spec}}$} &
\colhead{ NUV\tablenotemark{a}}&
\colhead{$f_{\nu}{830}$\tablenotemark{b}} &
\colhead{$(f_{\nu}{1500}/f_{\nu}{830})_{\mathrm{obs}}$\tablenotemark{c}}&
\colhead{$f_{\mathrm{esc,rel}}$\tablenotemark{d}}&
\colhead{$f_{\mathrm{esc,rel}}$\tablenotemark{e}}}

\startdata
\tableline
C-UVLG-02  &149.94569 & 2.50422 &108& 0.667 &22.40 &2.880e-19&36.90    &$<$0.108  &$<$0.223\\
C-UVLG-05  &150.08705 & 2.30905 &108& 0.688 &22.60 &1.708e-17&118.94  &$<$0.034  &$<$0.069\\
C-UVLG-07  &150.21405 & 2.36754 &108& 0.669 &22.69 &9.144e-16&73.45    &$<$0.054  &$<$0.112\\
C-UVLG-08  &150.42432 & 2.03343 & 81&  0.685 &22.35 &7.305e-20&144.16  &$<$0.028  & $<$0.057\\
C-UVLG-11  &150.15372 & 1.84970 &108& 0.670 &22.39 &6.471e-17&112.37  &$<$0.036  & $<$0.073\\
C-UVLG-12  &149.85553 & 2.55685 &108& 0.665 &22.50 &1.054e-16&20.69    &$<$0.193  & $<$0.397\\
C-UVLG-15  &150.27368 & 2.55358 &108& 0.715\tablenotemark{e} &22.60 &3.243e-21&40.49    &$<$0.099  & $<$0.203\\
C-UVLG-18  &150.60126 & 2.71238 &108& 0.676 &22.53 &1.671e-16&34.02    &$<$0.118  & $<$0.242\\
C-UVLG-19  &150.48743 & 2.15107 &108& 0.684 &22.41 &4.484e-20&34.73    &$<$0.115  & $<$0.237\\
C-UVLG-20  &150.51077 & 2.75674 &  81& 0.675 &22.09 &5.094e-19&9312    &$<$0.043  & $<$0.088\\
C-UVLG-21  &149.62891 & 2.19084 &108& 0.701 &22.46 &2.034e-16&161.10  &$<$0.025  & $<$0.051\\
C-UVLG-23  &149.88934 & 2.73473 &  27& 0.709 &21.41 &8.128e-19&203.56  &$<$0.019  & $<$0.040\\
C-UVLG-24  &150.17247 & 2.63495 &108& 0.752 &22.43 &1.660e-16&36.05    &$<$0.111  & $<$0.228\\
C-UVLG-25  &149.62904 & 1.66577 &  81& 0.671 &22.02 &1.182e-20&74.98    &$<$0.053  & $<$0.109\\
C-UVLG-26  &149.60760 & 2.16737 &108& 0.676 &22.28 &6.302e-19&121.33  &$<$0.033  & $<$0.068\\
C-UVLG-27  &149.69514 & 2.74294 &108&  0.652&22.52 &1.015e-19&82.99    &$<$0.048  & $<$0.099\\
C-UVLG-28  &150.45580 & 1.65284 &  81& 0.657 &22.66 &4.344e-19&33.84    &$<$0.118  & $<$0.243\\
C-UVLG-29  &149.46870 & 2.58616 &  81& 0.704 &21.94 &1.005e-18&66.43    &$<$0.060  & $<$0.124\\
C-UVLG-30  &149.51488 & 1.89228 &108& 0.694 &22.56 &8.146e-17&50.50    &$<$0.079  & $<$0.163\\
C-UVLG-33  &150.76311 & 2.80412 &108& 0.693 &22.37 &3.299e-19&124.97  &$<$0.032  & $<$0.065\\
C-UVLG-34  &150.69376 & 2.84451 &108& 0.738 &22.68 &8.607e-17&62.97    &$<$0.063  & $<$0.130\\
C-UVLG-35  &149.69348 & 2.59921 &  27& 0.678 &21.35 &1.490e-18&72.07    &$<$0.055   & $<$0.114\\
\hline
Stack &(18 galaxies)\tablenotemark{f}& & & 0.685 & 22.14 &6.379e-19&378.7 & $<$0.011 & $<$0.022 \\
Stack &(8  mergers)\tablenotemark{g}& & & 0.678 & 22.50 &1.249e-18&223.2 &$<$0.018 & $<$0.037 \\
\hline\hline

\enddata
\tablecomments{The values presents are $3\sigma$ limits. Objects 3, 9, 10, 13, 14, 16, 22, 31, and 32 were not detected in the FUV direct image, target 17 was an AGN.  }
\tablenotetext{a}{NUV magnitude from {\it GALEX} public release ver. 3}
\tablenotetext{b}{3$\sigma$ upper limits in units of erg s$^{-1}$cm$^{-2}$\AA$^{-1}$}
\tablenotetext{c}{3$\sigma$ lower limits}
\tablenotetext{d}{Assumes and intrinsic flux ratio of $(f_{1500}/f_{830})_{\mathrm{int}}=3.4$ based in part from \citep{2001ApJ...546..665S}}
\tablenotetext{e}{Assumes $(f_{1500}/f_{830})_{\mathrm{int}}=7$ \citep{2007ApJ...668...62S}}
\tablenotetext{f}{Photometric redshift}
\tablenotetext{g}{All objects except 2, 7, 15, and 18 are included in the stack due to their extended sizes} 
\tablenotetext{h}{Mergers that were compact with optimal dispersion directions: objects 5, 12, 19, 25, 26, 28, 30, and 35} 

\end{deluxetable}
\end{center}

\begin{figure}[h]
  \centering
  \includegraphics[width=170mm]{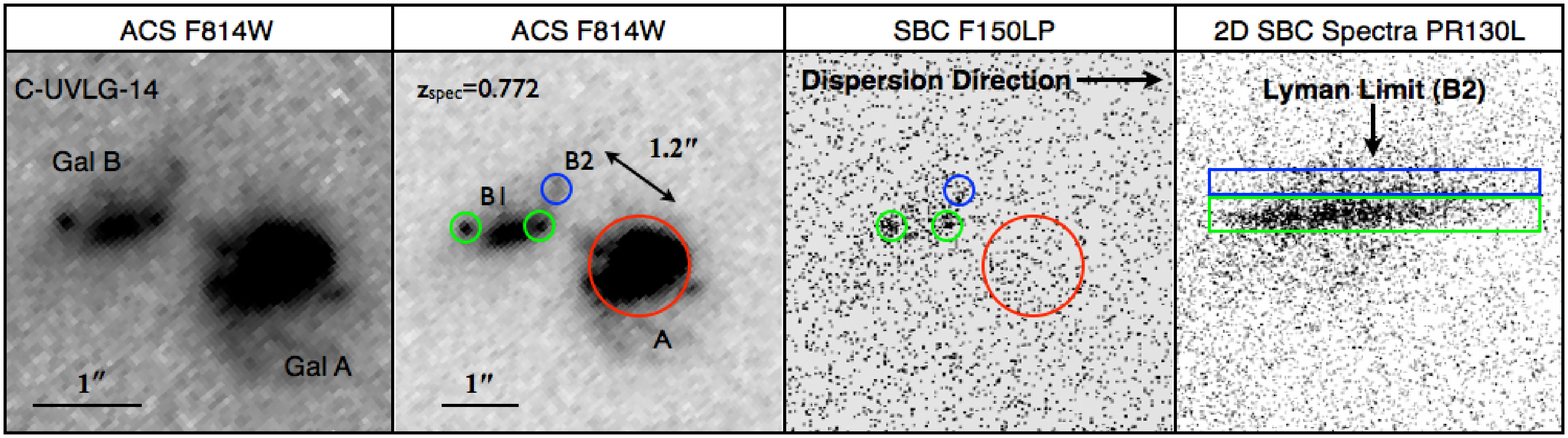}
\caption[LyC]{Postage stamp images left to right are \hst\, F814W zoomed in followed by F814W, SBC FUV F150LP, and the FUV 2D spectra. Galaxy A (GalA) and Galaxy B (GalB) are noted in the images.  The FUV flux and sites of escaping LyC are concentrated in knots (highlighted by circles) in GalB and are referred to in the text as GalB1 and GalB2.  The merging pair is separated by $\sim$10kpc on the sky. Boxes approximate the extraction regions of the UV spectra in the two-dimensional spectral image.  Flux below the Lyman limit (assuming a $z=0.772$) is seen to the right of the arrow.}
\label{fig:panels}
\end{figure}

\begin{figure}[h]
  \centering
  \includegraphics[width=130mm]{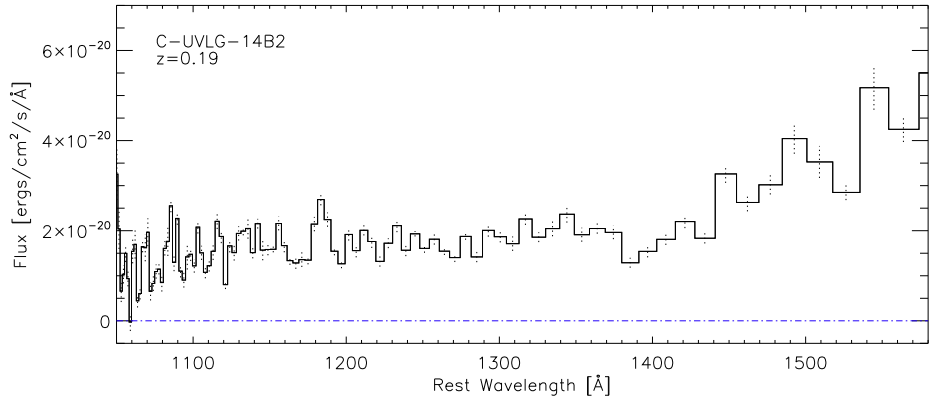}
\caption[uvspectra]{One-dimensional rest-frame UV spectra of GalB2, a low-redshift interloper at $z=0.19$.}
\label{fig:interloper}
\end{figure}

\begin{figure}
  \centering
  \includegraphics[width=65mm]{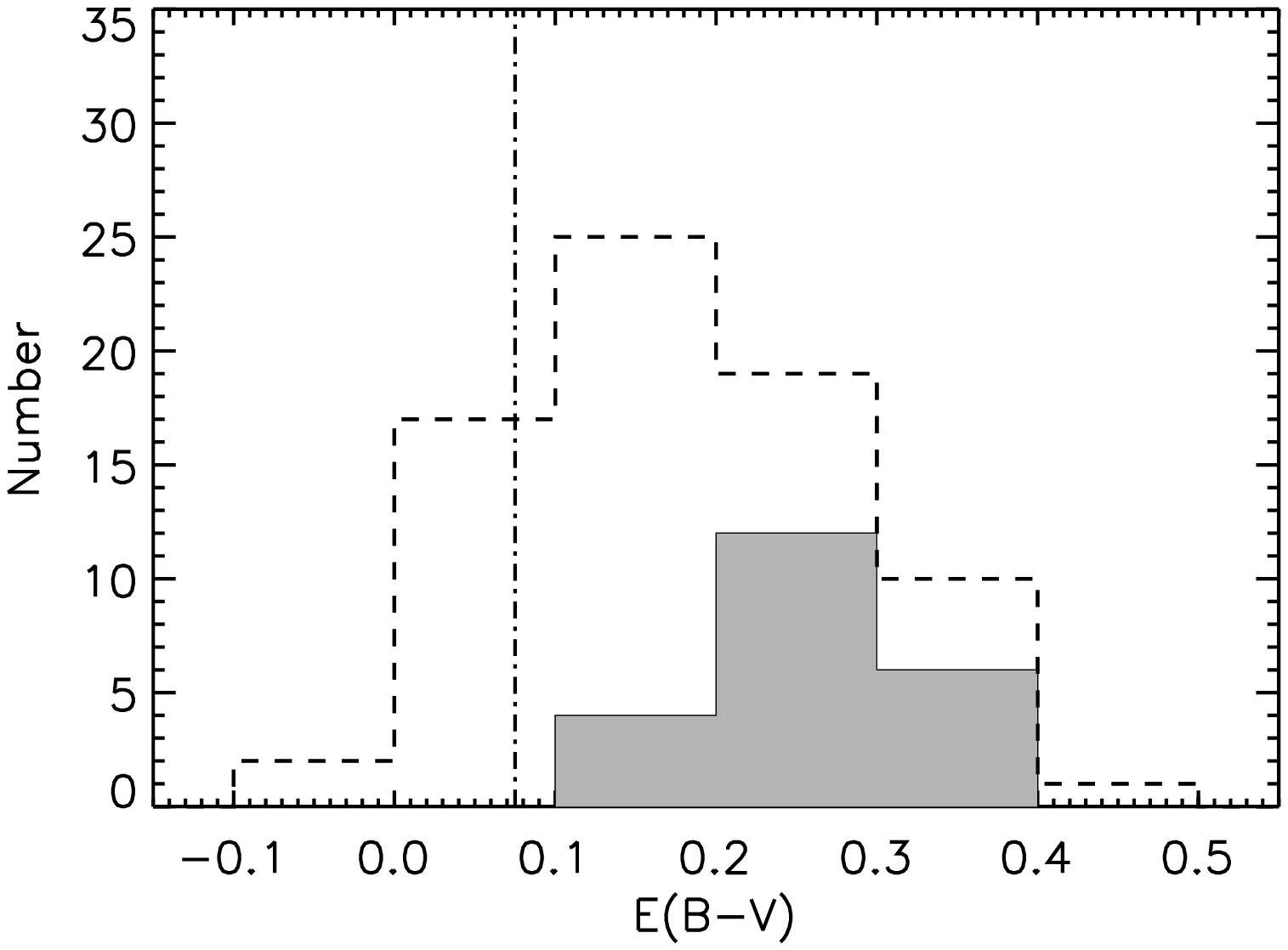}
  \includegraphics[width=65mm]{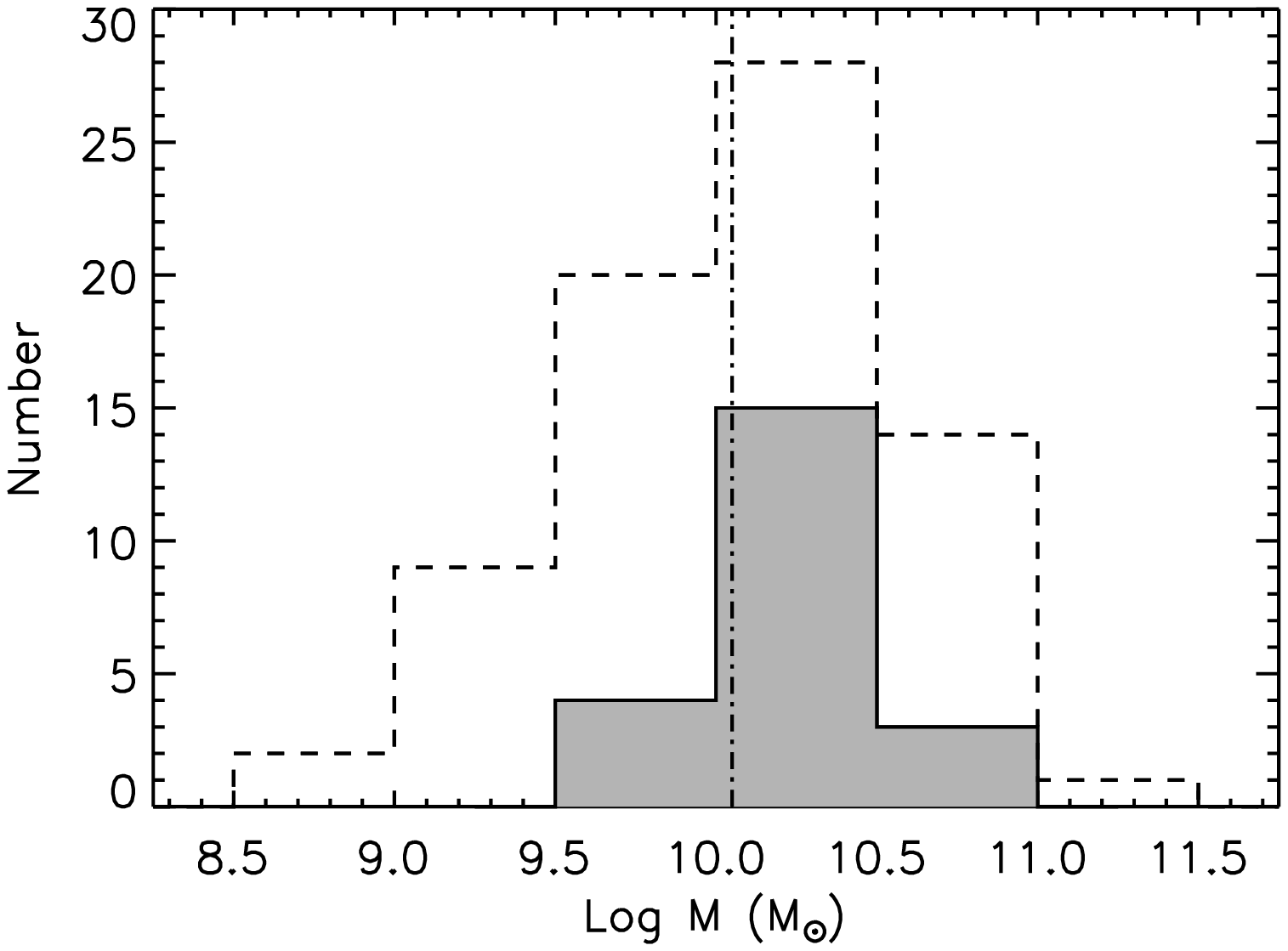}
\includegraphics[width=65mm]{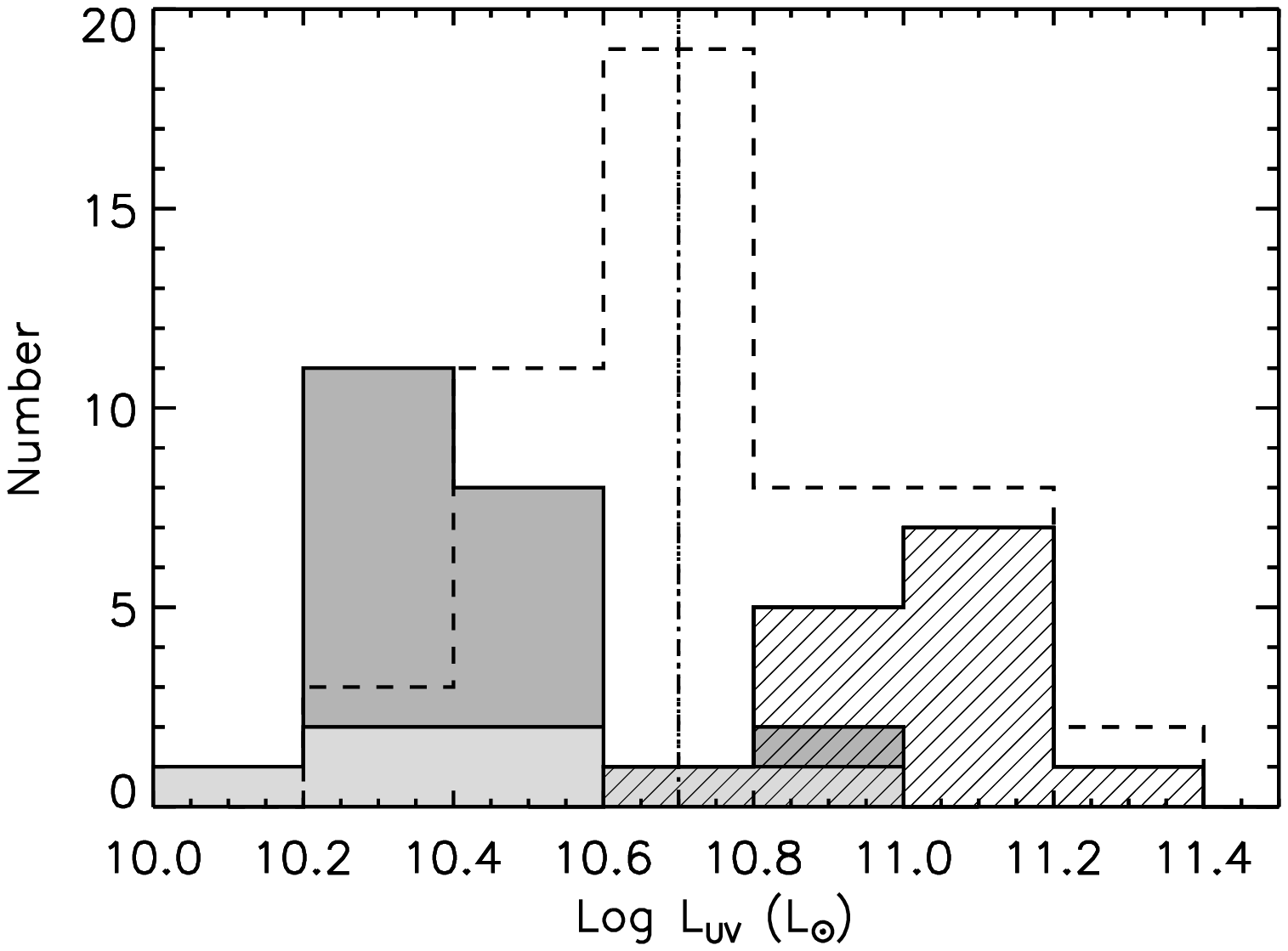}
\caption[ebv_mass]{Histograms of extinction (left), stellar mass (right) and rest-frame UV luminosity (bottom) for the
  22 local LBG analogs with measurable escape fractions. The darker shaded regions highlight the 22 galaxies presented in Table 1.  A sample of $z\sim3$ LBGs (dashed
  lines) in all three panels are taken from  \cite{2005ApJ...626..698S}.  The dash-dotted vertical lines show the average extinction \citep{2007MNRAS.377.1024V}, and stellar mass \citep{2006ApJ...651...24Y} for $z\gsim5$ galaxies and the typical $L_{*}$ for $z\sim6$ galaxies \citep{2006ApJ...653...53B}. (Bottom) The seven direct detections of LyC flux in LBGs reported in \citet{2008arXiv0805.4012I} are denoted by the lighter shaded histogram, and the lined (angled) histogram shows the $z\sim3$ LBG sample of \citet{2006ApJ...651..688S}.  The extinction and stellar mass for the COSMOS sources were estimated in the photometric redshift fitting \citep{2007ApJS..172..117M}, and the UV luminosity from the \galex\, NUV \citep{2007ApJS..172..468Z}. The histograms show that the galaxies presented in this paper have low levels of extinction, similar stellar masses and rest-frame UV luminosities compared to $z\sim3$ LBGs.}
\label{fig:ebv_mass}
\end{figure}

\end{document}